\documentclass[11pt, prb]{revtex4-1}

\usepackage{graphicx,rotating,subfigure}
\usepackage{amsmath}
\usepackage{pstricks}
\begin{document}

\title{Cooee bitumen II: Stability of linear asphaltene nanoaggregates}

\author{Claire A. Lemarchand}
\email{clairel@ruc.dk}
\affiliation{
  DNRF Centre ``Glass and Time'', IMFUFA,
  Department of Sciences, 
  Roskilde University, Postbox 260,
  DK-4000 Roskilde, Denmark $\mbox{}$
}

\author{Thomas B. Schr{\o}der}
\affiliation{
  DNRF Centre ``Glass and Time'', IMFUFA,
  Department of Sciences,
  Roskilde University, Postbox 260,
  DK-4000 Roskilde, Denmark $\mbox{} $ $\mbox{} \mbox{}$
}

\author{Jeppe C. Dyre}
\affiliation{
  DNRF Centre ``Glass and Time'', IMFUFA,
  Department of Sciences, 
  Roskilde University, Postbox 260,
  DK-4000 Roskilde, Denmark $\mbox{} $ $\mbox{}$
}

\author{Jesper S. Hansen}
\affiliation{
  DNRF Centre ``Glass and Time'', IMFUFA,
  Department of Sciences,
  Roskilde University, Postbox 260,
  DK-4000 Roskilde, Denmark
}

\begin{abstract}
Asphaltene and smaller aromatic molecules tend to form linear nanoaggregates in bitumen.
Over the years
bitumen undergoes chemical aging and during this process, the size of the nanoaggregate increases. 
This increase is associated with an increase in viscosity and brittleness of the bitumen,
eventually leading to road deterioration. 
This paper focuses on understanding the mechanisms behind nanoaggregate size and stability.
We used molecular dynamics
simulations to quantify the probability of having a nanoaggregate of a given size in the stationary regime. 
To model this complicated behavior, we chose first to consider the simple case where only asphaltene molecules
are counted in a nanoaggregate.
We used a master equation approach and a related statistical mechanics model.
The linear asphaltene nanoaggregates behave as a rigid linear chain.
The most complicated case where all aromatic molecules are counted in a nanoaggregate is then discussed.
The linear aggregates where all aromatic molecules are counted seem to behave as a flexible linear chain.
\end{abstract}

\maketitle

\noindent Corresponding author: C. A. Lemarchand, E-mail: clairel@ruc.dk

\newpage \setlength{\baselineskip}{0.75cm}

\section{Introduction}
One of the main industrial applications of bitumen is as a binder in asphalt pavement~[\onlinecite{ec140}].
Bitumen links together mineral aggregates and filler particles to form
a cohesive asphalt on the road surface.
Over the years, a chemical reaction takes place in bitumen increasing
the number of heavy aromatic molecules~[\onlinecite{ec140, shrp368, herrington, lu, petersen74}].
This process is called chemical aging.
The aromatic molecules in bitumen, especially the asphaltene molecules, tend to align to form nanoaggregates.
The aging reaction leads also to an increase
in the nanoaggregate size~[\onlinecite{ec140, mullins2012, aging}] correlated with an unwanted increase in bitumen viscosity and
brittleness~[\onlinecite{shrp368, herrington, lu, ec140, petersen74}].
The change in bitumen mechanical properties, which go from "liquid-like" to more
"solid-like" during chemical aging finally results in cracks in the pavement and road deterioration. 
To prevent or reverse the effects of chemical aging, a first step is to gather more knowledge about the 
nanoaggregate structure and stability in bitumen.

Much progress has been made over the last 50 years in the experimental and numerical literature
to determine the structure of the nanoaggregates and the conditions under which they are formed.
One can cite in particular the design of the Yen-Mullins model~[\onlinecite{yen}, \onlinecite{mullins2012}] describing
the nanoaggregate structure, the
determination of the critical nanoaggregate concentration in
different solvents~[\onlinecite{andreatta}, \onlinecite{goual}],
the evaluation of the average size and polydispersity of a nanoaggregate~[\onlinecite{eyssautier2011}] on
the experimental side.
On the numerical side, different stable conformations of nanoaggregates were identified
depending on the asphaltene structure using molecular
mechanical calculations~[\onlinecite{murgich1996}] and molecular dynamics (MD) simulations~[\onlinecite{pacheco2004}].
MD simulations were also used to
determine the molecular orientation inside the nanoaggregates~[\onlinecite{zhang2007}],
and the effects of solvent and
presence of other molecules on the nanoaggregate structure~[\onlinecite{headen2011}].
However, analytical models for the thermodynamics stability and dynamics
of the nanoaggregates in relation to their structure are still quite rare, to the notable exception
of Ref.~[\onlinecite{aguilera2006}]. The purpose of this
paper is precisely to suggest simple and generic models,
which can reproduce MD results on the nanoaggregate stability.

We present molecular dynamics results
concerning the nanoaggregate size in the stationary regime.
In bitumen, nanoaggregates are composed of asphaltene molecules, the most heavy and aromatic fraction in bitumen,
but also of lighter aromatic molecules like resin and resinous oil~[\onlinecite{aging}]. We give
results on 
the probability of having a nanoaggregate containing a given number of aromatic molecules and the probability
of having a nanoaggregate containing a given number of asphaltene molecules.
The two probabilities are shown to differ qualitatively.
We first model the simpler case, where only asphaltene
molecules are counted.
The focus in this case is to establish a mechanism for the nanoaggregation 
of asphaltene molecules. From this aggregation mechanism,
a master equation and a
statistical mechanics model are derived and provide a simple theoretical framework
for interpreting the simulation results on nanoaggregate stability.
The results of the master equation approach are also compared to the aggregation dynamics observed
in the MD simulations.
Then, the more
complicated case where all aromatic molecules are counted is discussed
in terms of statistical mechanics arguments.

The MD simulations carried out in this work are based on the four-component united-atom-unit model
developed in Ref.~[\onlinecite{us}] in the framework of the COOEE project~[\onlinecite{cooee}]. The
model is shown to reproduce a generic bitumen reasonably well~[\onlinecite{us}].
The simulations were performed on Graphic-Processor-Units (GPU) using
the Roskilde University Molecular Dynamics (RUMD.org) package~[\onlinecite{rumd}].

The paper is organized as follows. In Sec.~\ref{sec:results}, 
we provide simulation details, give a definition of
a linear nanoaggregate, and present the MD results about the probability of having
a nanoaggregate of a given size. Section~\ref{sec:masterEq} is devoted to
model the results on the stability of
linear aggregates where only asphaltene molecules are counted with a master equation approach. 
In Sec.~\ref{sec:statMech}, a related statistical mechanics model is described
and shown to reproduce the MD results on asphaltene nanoaggregate stability. The more complicated
case where all aromatic molecules are counted in a nanoaggregate is also discussed in Sec.~\ref{sec:statMech} 
in terms of statistical mechanics arguments.
Sec.~\ref{sec:discussion} includes a comparison of our results to existing experimental
and numerical results and discuss the limit of our model.
Finally, Sec.~\ref{sec:conclu} contains a summary and a conclusion.

\section{Molecular dynamics results}
\label{sec:results}

Before presenting the molecular dynamics (MD) results on nanoaggregate stability, we
mention a few details about the simulations and define precisely a nanoaggregate.

\subsection{Simulation details}
\label{sec:simulationDetail}
As mentioned in the introduction, the simulation method and molecular potentials are described in details in Ref.~[\onlinecite{us}].
Only information necessary to understand the study on nanoaggregate stability carried out in the present paper
is given here.

\begin{figure}
  \scalebox{0.20}{\includegraphics{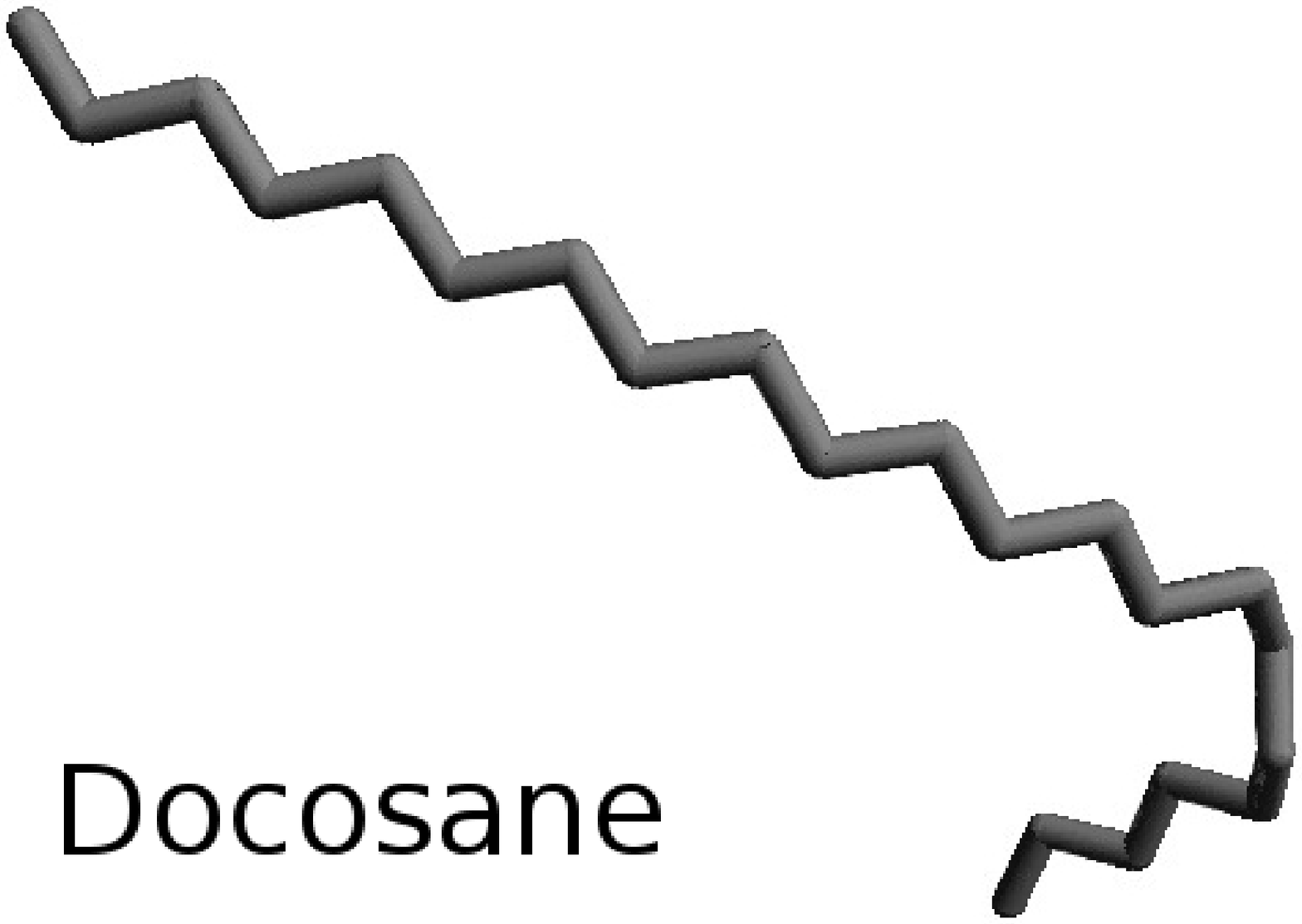}}
  \scalebox{0.20}{\includegraphics{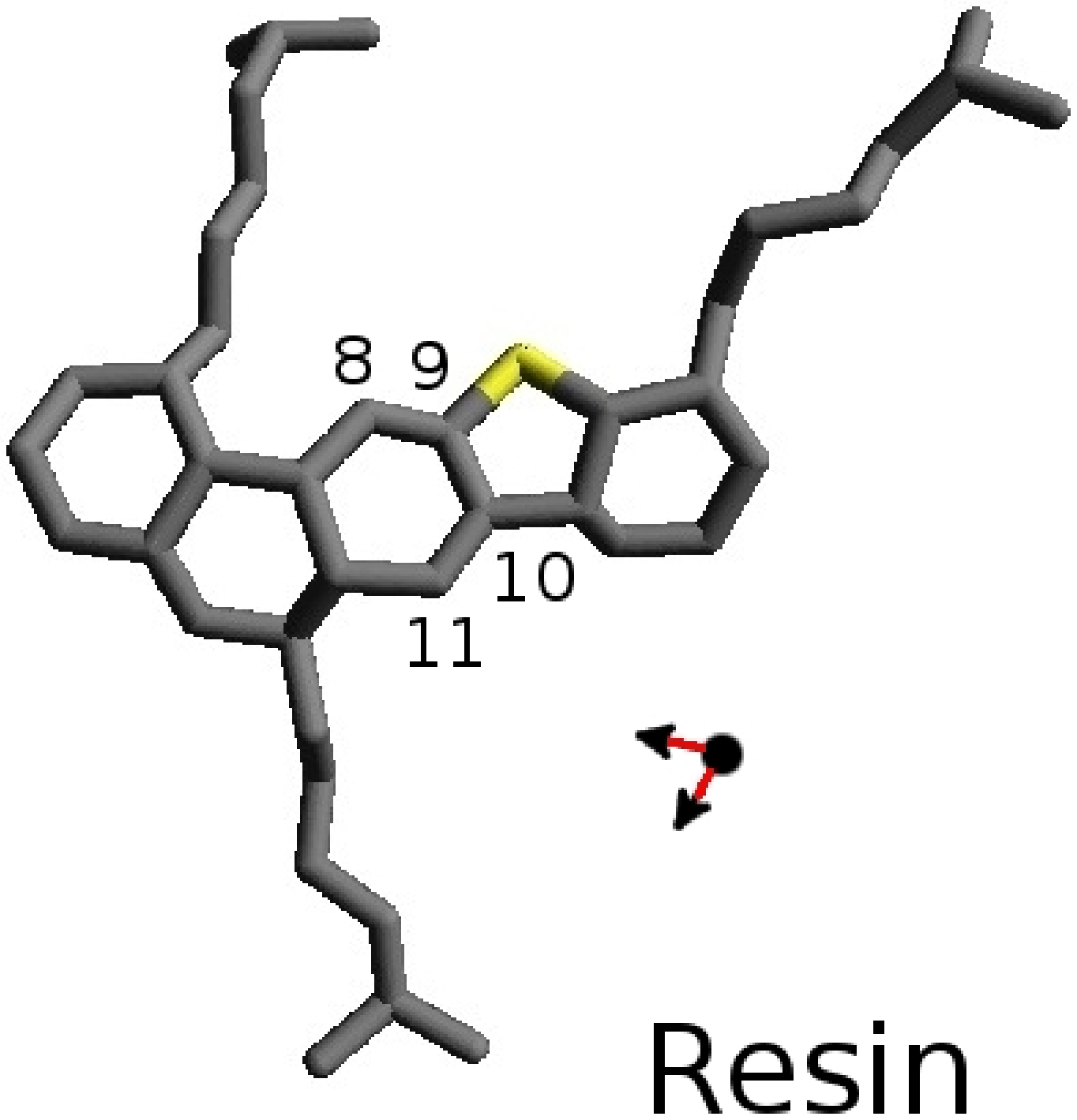}}\\
  \scalebox{0.25}{\includegraphics{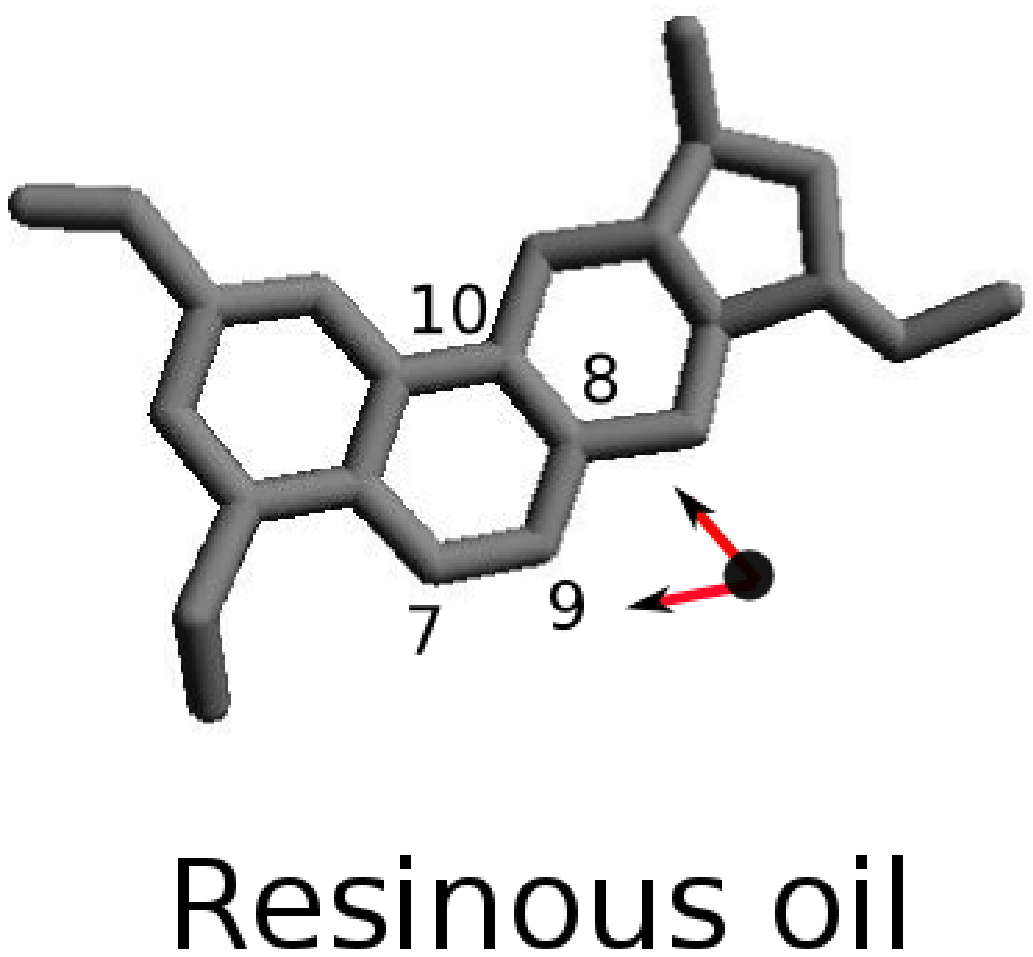}}
  \scalebox{0.25}{\includegraphics{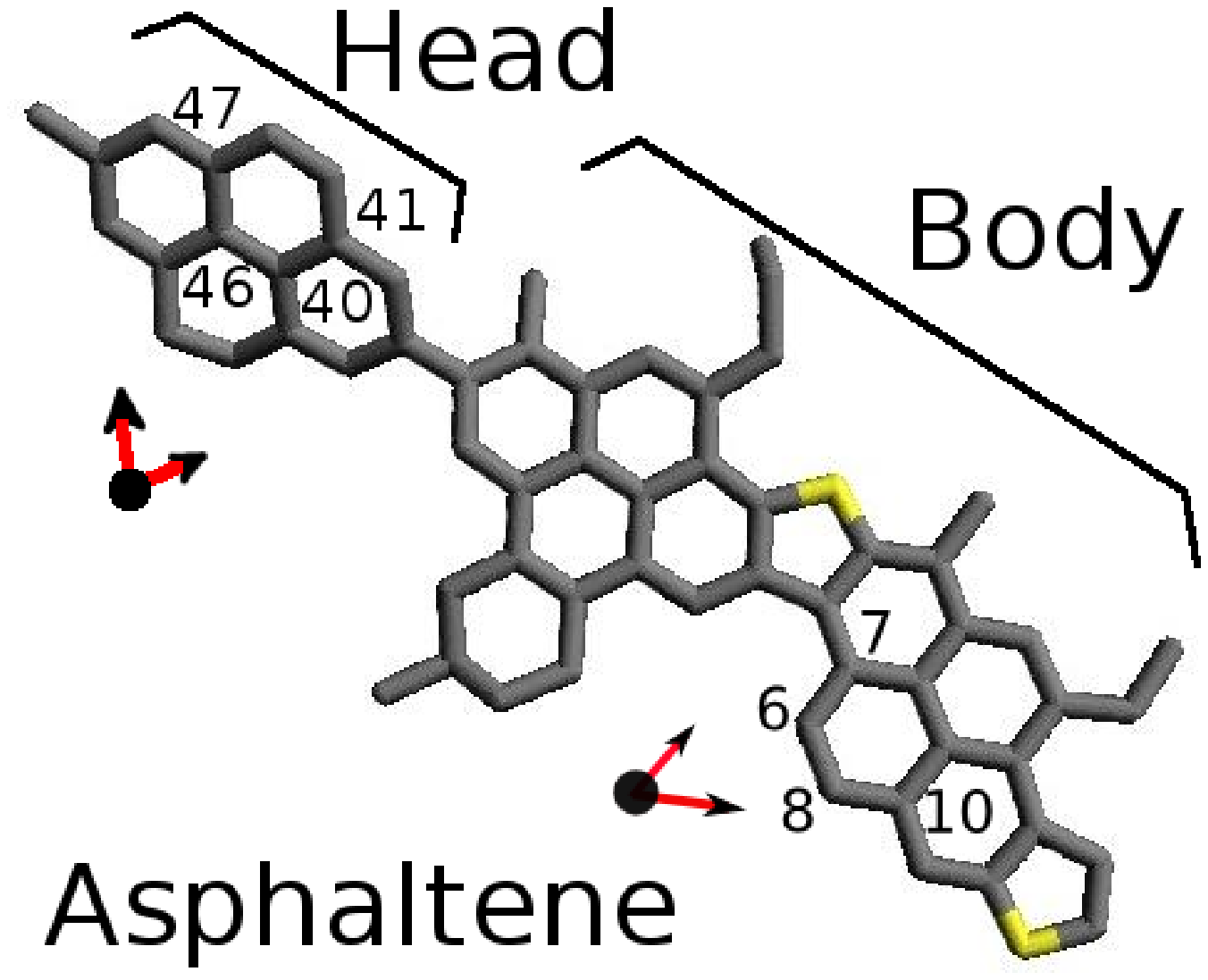}}
  \caption{\label{fig:molecule}
(Color online) Structure of the four molecules in the "COOEE bitumen" model. Grey edges represent
the carbon groups $CH_3$, $CH_2$ and $CH$ and yellow edges represent sulfur atoms.
%Top left: docosane. Top right: resin. Lower left: resinous oil. Lower right: asphaltene.
The "head" and "body" of the asphaltene molecule are shown.
Numbers and arrows indicate bond-vectors used to quantify the nanoaggregate structure.
Reprinted with permission from
C. A. Lemarchand, T. B. Schroder, J. C. Dyre, and J. S. Hansen, J. Chem. Phys. 139, 124506 (2013).
Copyright 2013 American Institute of Physics.
}
\end{figure}

The simulated system contains four types of molecule, chosen to resemble the SARA classification~[\onlinecite{SARA}]:
a Saturated hydrocarbon, a resinous oil molecule, which is denoted Aromatic in the SARA
scheme, a Resin molecule and an Asphaltene molecule.
The molecular structures chosen are shown in Fig.~\ref{fig:molecule}.
The main system studied in this paper contains $410$ saturated hydrocarbons, $50$ resinous oil molecules,
$50$ resin molecules and $50$ asphaltene molecules, which corresponds to $15570$ united atom units.
The methyl (CH$_3$), methylene (CH$_2$), and methine (CH) groups are represented by the
same united atom unit of molar mass 13.3 g$\cdot$mol$^{-1}$ and the sulfur atoms are represented
by a united atom unit with a different molar mass 32 g$\cdot$mol$^{-1}$.
The potential between the united atom units contains four terms: an intermolecular potential,
corresponding to a Lennard-Jones potential with parameters $\sigma = 3.75$ \AA $ $ and $\epsilon/k_B = 75.4$ K,
where $k_B$ is the Boltzmann constant and three terms for the intramolecular potential.
These three terms describe the bond length between two connected particles,
the angle between three consecutive particles, and the dihedral angle between four consecutive particles.
The parametrization of the intramolecular potential is described in details in Refs~[\onlinecite{us}].
The simulations are performed in the canonical ensemble (NVT)
at a constant temperature $T = 452$ K and a constant density. The density $\rho = 0.964$ kg$\cdot$L$^{-1}$ is chosen to obtain
an average pressure around the atmospheric pressure. A Nos\'e-Hoover thermostat is used.
The time step is $\Delta t = 0.86$ fs and the duration of the simulations is $T = 360$ ns.
Eight independent simulations are performed
at the same state point. The molecular dynamics package RUMD~[\onlinecite{rumd}] is used to perform the calculation.

\subsection{Definition of a nanoaggregate}
\label{sec:defNanoaggregate}
It was shown in a previous work~[\onlinecite{aging}] that the asphaltene, resin, and resinous oil molecules
tend to align with respect to each other. They align at a distance of around $4.0$ \AA, close to the
minimum of the Lennard-Jones potential between the molecules. 
The Lennard-Jones potential used in the MD simulations mimics the $\pi$-stacking interaction observed between
aromatic molecules experimentally.
The alignment of aromatic molecules in bitumen is the basis of the nanoaggregate
formation.

\begin{figure}
  \includegraphics[scale=0.7]{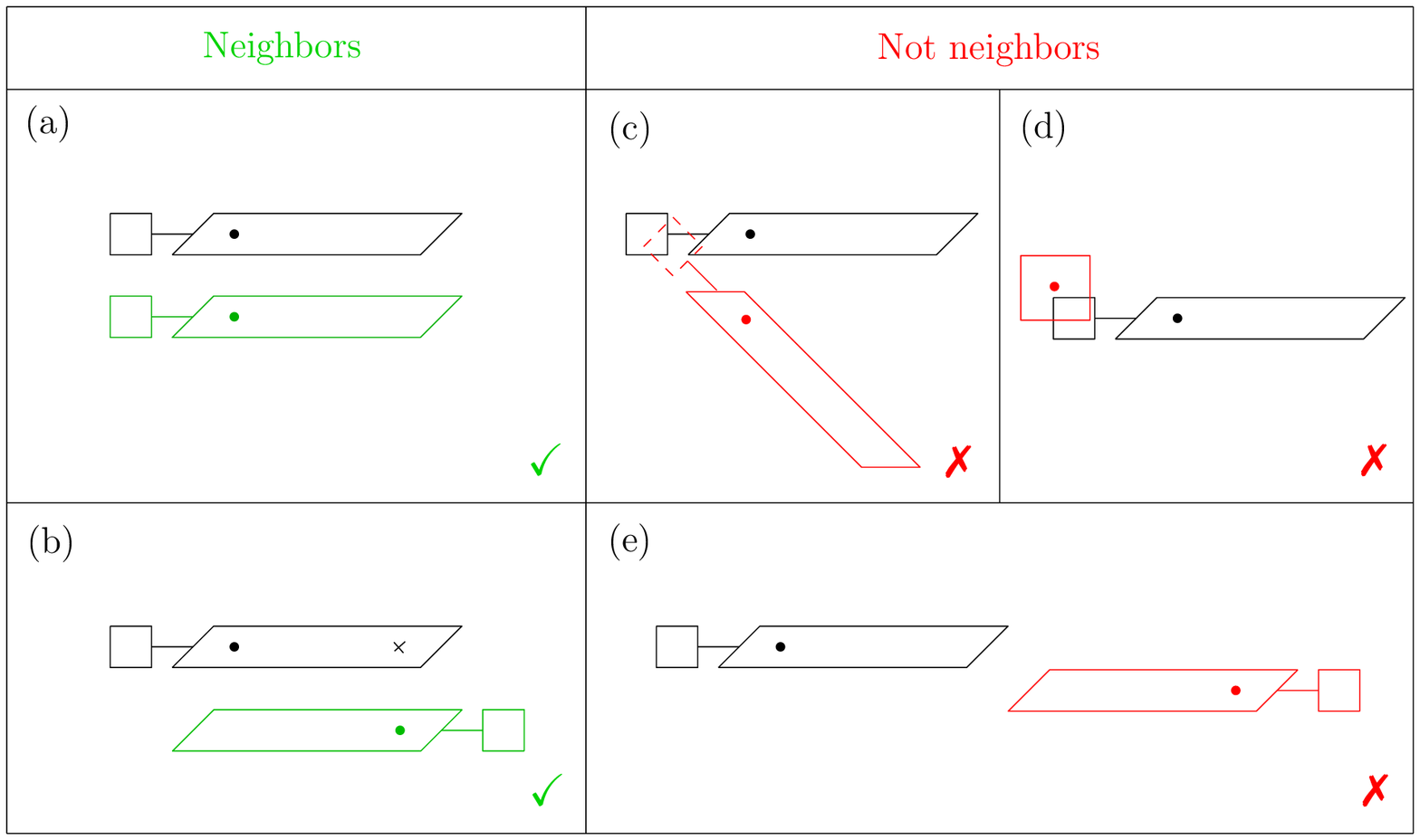}
  \includegraphics[scale=0.35, clip=true, trim=0cm 6cm 0cm 6cm]{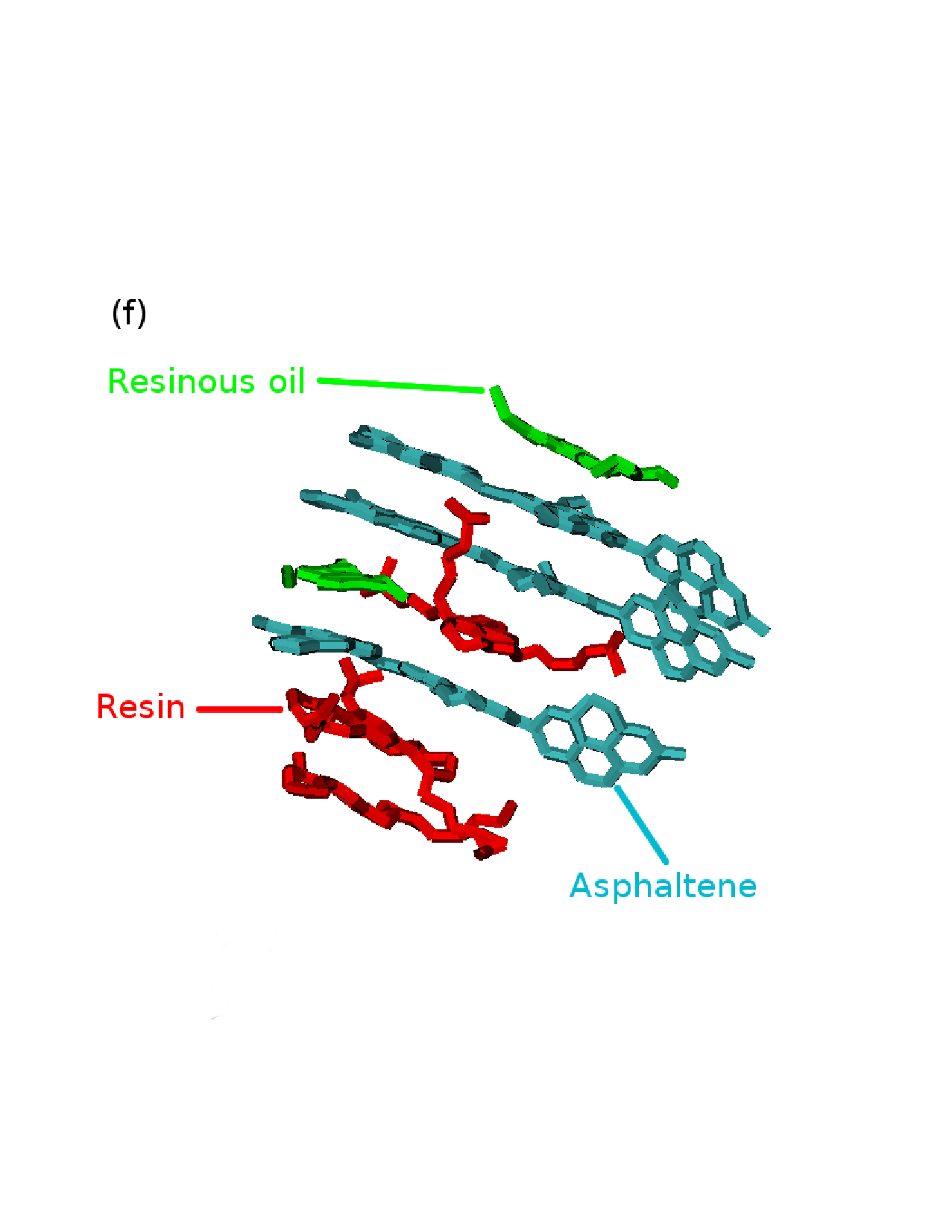}
  \caption{\label{fig:linearAggregate}
(Color online) (a) to (e) Scheme of limiting cases illustrating whether a
molecule is or is not the nearest neighbor of the first molecule.
The first molecule is an asphaltene molecule and is black.
The second molecule is green if it is the neighbor of the first one and red
otherwise.
(a) Head-to-head conformation. (b) Head-to-tail conformation.
(c) Non-aligned molecules. (d) Molecule aligned to the head of the asphaltene
molecule. (e) Molecules far from each other and aligned.
Modified with permission from
C. A. Lemarchand, T. B. Schroder, J. C. Dyre, and J. S. Hansen, J. Chem. Phys. 139, 124506 (2013). 
Copyright 2013 American Institute of Physics.
(f) Snapshot of a linear nanoaggregate, obtained in molecular dynamics. Asphaltene molecules
are in blue, resin molecules in red and resinous oil molecules in green.
}

\end{figure}

The definition of a nanoaggregate is described in detail in Ref.~[\onlinecite{aging}]. It is based on the following
rule: two aromatic molecules are nearest neighbors in the same nanoaggregate if they are "well-aligned"
and "close enough".
A nanoaggregate is composed of all molecules connected by this rule.
Moreover, the asphaltene molecule chosen in this model has two parts,
a flat head and a flat body which can rotate with respect to each other.
For this reason, aromatic molecules can align
in the direction of an asphaltene body or in the direction of an asphaltene head,
thus creating branched nanoaggregates.
These branches can link together purely linear nanoaggregates.
We believe this is one mechanism explaining the formation of clusters of nanoaggregates, also observed
experimentally~[\onlinecite{mullins2012}].
In this paper, we focus on purely linear nanoaggregates. 
It means that any aromatic molecules
linked to an asphaltene head will not be considered as part of this asphaltene nanoaggregate.
We  define a linear aggregate as composed of asphaltene bodies, resin and resinous oil molecules,
and not asphaltene heads
because heads are the smallest parts and probably lead to the smallest
interaction energy.
Figure~\ref{fig:linearAggregate} shows conformations
of two molecules corresponding to limiting cases of the nanoaggregate definition
used in this paper.
Figures~\ref{fig:linearAggregate}(a) and (b) show two conformations
of two asphaltene molecules where these molecules are considered as nearest neighbors.
Conversely, Figs~\ref{fig:linearAggregate}(c) to (e) show three conformations
where the two molecules are not considered as nearest neighbors.
A picture of a linear nanoaggregate obtained
from the MD simulations is shown in Fig.~\ref{fig:linearAggregate}(f).
Note that this nanoaggregate is not typical as a resin and a resinous oil molecule
are aligned on the same side of an asphaltene body.

\subsection{Probability of having a nanoaggregate of a given size}
\label{sec:MDresults}
Sections~\ref{sec:simulationDetail} and~\ref{sec:defNanoaggregate} gave necessary information about
the simulations and the definition of linear nanoaggregates. The results of the molecular
dynamics simulations concerning the nanoaggregate size and stability can now be presented.

There are different possibilities to quantify the size of a linear nanoaggregate.
For example, it can be quantified as the total number of aromatic molecules or as the number of asphaltene
molecules which it contains. The first definition makes use of the fact that nanoaggregates are not only composed
of asphaltene molecules, but also of smaller aromatic molecules. 
The second definition accounts for the fact that asphaltene
molecules are the largest aromatic molecules and probably the most important in the nanoaggregate stability.
This is why experimentalists often define the nanoaggregate size as the latter~[\onlinecite{mullins2012}].

\begin{figure}
  \includegraphics[angle=-90, scale=0.4]{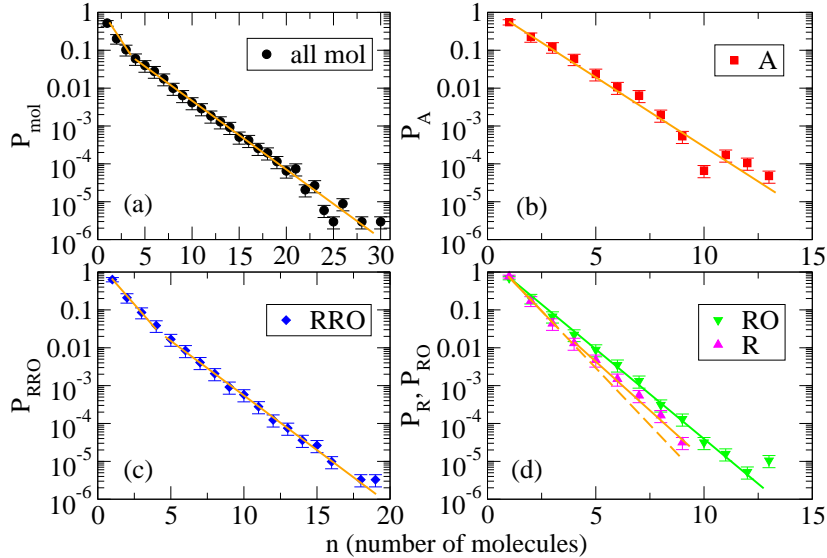}
  \caption{\label{fig:pdfall}
(Color online) (a) Probability $P_{\text{mol}}(n)$ of having a nanoaggregate containing $n$ aromatic molecules versus $n$.
(b) Probability $P_A(n)$ of having a nanoaggregate containing $n$ asphaltene molecules versus $n$.
(c) Probability $P_{RRO}(n)$ of having a nanoaggregate containing $n$ resin or resinous oil molecules versus $n$.
(d) Probability $P_R(n)$ of having a nanoaggregate containing $n$ resin molecules versus $n$ and probability $P_{RO}$
of having a nanoaggregate containing $n$ resinous oil molecules versus $n$.
In every case the straight lines (orange or green) are guides to the eye. They highlight
the monoexponential behavior in the subfigures (b) and (d), when resinous oil molecules are counted,
and the biexponential behavior in the subfigures (a), (c), and (d), when resin molecules are counted.
The error bars
correspond to the standard deviation estimated from eight independent simulations.
}

\end{figure}

The molecular dynamics simulations enable us to study the consequences of both definitions
on the nanoaggregate stability. We quantify the stability as the probability of having a nanoaggregate
of a given size in the stationary regime. 
The probability $P_{\text{mol}}(n)$ of having a nanoaggregate containing $n$ aromatic molecules is defined as:
\begin{equation}
P_{\text{mol}}(n) = \frac{N_{n, \text{mol}}}{N_{t}},
\end{equation}
where $N_{n, \text{mol}}$ is the number of linear aggregates containing $n$ aromatic molecules and $N_{t}$ is the
total number of linear aggregates.
In a similar way, the probability $P_X(n)$ of having a nanoaggregate with $n$ molecules
of type $X$ is defined as
\begin{equation}
\label{eq:PAdef}
P_X(n) = \frac{N_{n, X}}{N_{t, X}},
\end{equation}
where $N_{n, X}$ is the number of linear aggregates containing $n$ molecules of type $X$
and $N_{t, X}$ is the
total number of linear aggregates containing at least one molecule of type $X$.
The type $X$ can be $A$ for asphaltene, $R$ for resin, $RO$ for resinous oil or $RRO$ for resin and resinous oil.
The probabilities $P_A(n)$ and $P_{RRO}(n)$ are plotted in Fig.~\ref{fig:pdfall} (b) and (c), respectively.
The probabilities $P_{R}(n)$ and $P_{RO}(n)$ are
plotted in Fig.~\ref{fig:pdfall} (d).
Two main points can be noticed in Fig.~\ref{fig:pdfall}. The probability $P_{\text{mol}}$ of having a nanoaggregate containing $n$ aromatic
molecules seems to have two slopes in a log-lin scale, \emph{i. e.} can be described by a biexponential. On the contrary,
when only asphaltene molecules are counted, the probability $P_A$ has only one slope, \emph{i. e.} 
can be characterized as a simple exponential.
For the probability $P_{RRO}$ of having a nanoaggregate containing $n$ resin or resinous oil molecules, the biexponential
shape seems to be prevailing. It is the same for the probability $P_R$ of having a nanoaggregate containing $n$ resin molecules.
For the probability $P_{RO}$ of having a nanoaggregate containing $n$ resinous oil molecules, the monoexponential
shape seems to be recovered.
The transition between the two slopes is not sharp for the probability $P_R$. To highlight the existence
of the two slopes in this case in Fig.~\ref{fig:pdfall} (d), a dashed line was drawn in continuation of
the line corresponding to the first slope. It departs further and further away from the second slope.

The rest of the paper is devoted to understand the molecular dynamics results presented in Fig.~\ref{fig:pdfall}.
More specifically, the paper is aimed at unraveling some of the dynamical and thermodynamical
origins lying behind the monoexponential and biexponential behaviors. 
A dynamical approach will first be proposed for the monoexponential behavior.

\section{Dynamical approach}
\label{sec:masterEq}
In Fig.~\ref{fig:pdfall} (b), the probability of having a nanoaggregate containing $n$ asphaltene molecules
was shown to be monoexponential. The aim of this section is to establish a simple aggregation mechanism
reproducing this behavior. The mechanism will be used to derive a master equation
which provides
a dynamical 
framework for understanding the monoexponential behavior.

A simple mechanism
which can be written in the asphaltene aggregation problem is
\begin{equation}
\label{eq:reaction}
A_n +A_1 \rightleftharpoons A_{n+1}, \quad n = 1, 2, ...
\end{equation}
where $A_n$ denotes an aggregate containing $n$ asphaltene molecules.
The assumptions lying behind this mechanism are the following:
only one asphaltene
molecule at a time can attach to or detach from an existing aggregate, the medium is
homogeneous and the aggregation dynamics is not limited by diffusion.
The same mechanism was already suggested in Ref.~[\onlinecite{aguilera2006}] to
model asphaltene aggregation from a thermodynamical point of view,
but the possibility to obtain a master equation from this mechanism
and consequently to model the dynamics of asphaltene
aggregation was left aside.

To obtain a master equation linked to this aggregation reaction, we
make supplementary assumptions:
the reaction rate constants do not depend on the size $n$ of the aggregate and
the attachment rate does not depend on the probability of having free asphaltene molecules
in the medium.
Under these assumptions, the master equation
linked to the aggregation reaction~Eq.~\eqref{eq:reaction} can be written as:
\begin{align}
\label{eq:BC}
\frac{d P_A(1,t|n_0)}{dt} &= \mu P_A(2,t|n_0) - (\lambda+\mu) P_A(1,t|n_0) +\mu-\lambda, \\
\label{eq:masterEq}
\frac{d P_A(n,t|n_0)}{dt} &= \lambda P_A(n-1,t|n_0) + \mu P_A(n+1,t|n_0) - (\lambda+\mu)P_A(n,t|n_0), \quad \text{for} \quad n = 2, 3, ...
\end{align}
where $\lambda$ is the attachment rate constant, $\mu$ the detachment rate constant and $P_A(n,t|n_0)$ the probability
of having an aggregate containing $n$ asphaltene molecules at time $t$, given that its initial size was $n_0$.
The probability denoted $P_A(n)$ in Sec.~\ref{sec:MDresults} corresponds to the stationary state 
to Eqs.~\eqref{eq:BC} and~\eqref{eq:masterEq} and depends
neither on time $t$ nor on the initial condition $n_0$.
The master equation Eq.~\eqref{eq:masterEq}
was solved analytically in the 1950's~[\onlinecite{morse55},\onlinecite{schwarz75}] for different boundary conditions.
This master equation and its generalized version in which the rate constants depend on the aggregate size $n$
have also been extensively used to model different stochastic processes such as
alcohol clusters~[\onlinecite{sillrenMasterEq}, \onlinecite{thesisPer}], biological adhesion clusters~[\onlinecite{erdmann}] or aggregation
in freeway traffic~[\onlinecite{mahnke}], to cite only a few.

The expression of this master equation deserves some further clarification in connection to the specific aggregation reaction used.
Following the law of chemical kinetics, the attachment rate linked to the aggregation mechanism~\eqref{eq:reaction}
should be $\lambda  P_A(n-1,t|n_0) P_A(1,t|n_0)$, because the probability of having a collision between two small spherical molecules
is usually proportional to the product of their concentrations. 
We can give empirical reasons explaining why it is not proportional to
the probability $P_A(1,t|n_0)$ of having a free asphaltene molecule here.
We assume that the attachment rate is not due to the collision
between two small spherical molecules, but limited by the steric hindrance around the aggregate and the orientation
of the free asphaltene molecule $A_1$. The attachment reaction is successful if the 
free asphaltene molecule is well oriented and placed at a reactive end of
the aggregate. 
The number of free asphaltene molecules is higher than 
the number of any aggregate of a given size at any time in our MD simulations (not shown).
As the number of free asphaltene molecules is high enough,
there is always a free asphaltene molecule nearby an aggregate but it is not always well-oriented and prevent other molecules
from approaching due to the density of the system. Thus, the attachment rate is independent of the free asphaltene
concentration and the attachment rate
constant $\lambda$ is averaged over possible orientations.
The boundary condition~Eq.~\eqref{eq:BC} needs also some explanation. It corresponds to the fact that the system is closed.
In other words, the total number of asphaltene molecules is kept constant. 
As the free asphaltene molecules are involved in each aggregation reaction~\eqref{eq:reaction},
the derivative with respect to time of the probability $P(1,t|n_0)$ to have a free asphaltene molecule
depends on infinitely many terms. 
It leads to
\begin{equation}
\frac{d P_A(1,t|n_0)}{dt} = 2  \mu P_A(2,t|n_0) - 2 \lambda  P_A(1,t|n_0) + \sum\limits_{n=3}^{\infty} \mu P_A(n,t|n_0) -  \sum\limits_{n=2}^{\infty} \lambda P_A(n,t|n_0).
\end{equation}
These series are then simplified using the
fact that at every time $t$:
$\Sigma_{n=1} P_A(n,t|n_0) = 1$, to give the boundary condition~Eq.~\eqref{eq:BC}.

To test the validity of this master equation for the present problem,
two different quantities are evaluated : 
the stationary probability of having a nanoaggregate of a given size, as shown in Sec.~\ref{sec:MDresults}
and the aggregation dynamics, quantified as the time evolution of the fraction of aggregated asphaltene molecules.

\subsection{Stationary regime}
\begin{figure}
  \includegraphics[angle=-90, scale=0.4]{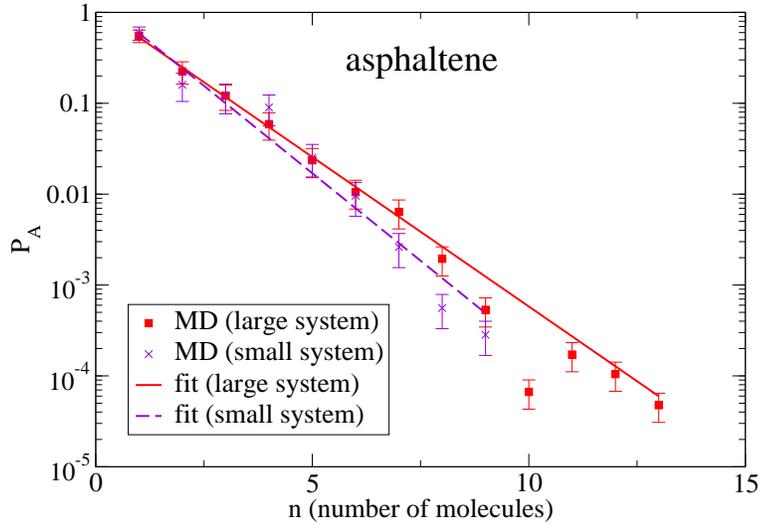}
  \caption{\label{fig:AonlyBB}
(Color online) Probability of having a nanoaggregate containing $n$ asphaltene molecules versus $n$.
Data for a large system, containing $50$ asphaltene molecules, $50$ resin molecules, $50$ resinous molecules
and $410$ docosane molecules and for a system $5$ times smaller are shown. The results for the large system
were already shown in Fig.~\ref{fig:pdfall} (b).
The red solid line is a fit of~Eq.~\eqref{eq:monoexpFit} to the data of the large system, with $p = 0.44$. 
The purple dashed line is a fit of~Eq.~\eqref{eq:monoexpFit} to the data of the small system, with $p = 0.41$.
}

\end{figure}

The stationary distribution of the master equation~Eq.~\eqref{eq:masterEq} with the boundary condition~Eq.~\eqref{eq:BC}
is easy to derive.
In the stationary state, the probability of having a nanoaggregate containing $n$ asphaltene molecules
depends neither on time nor on the initial size of a nanoaggregate.
For this reason, the stationary probability will be denoted $P_A(n)$. It satisfies the following equation:
\begin{align}
\label{eq:stationaryBC}
&(\lambda+\mu) P_A(1) = \mu P_A(2) + \mu -\lambda,\\
\label{eq:induction}
&\lambda P_A(n-1) + \mu P_A(n+1) - (\lambda+\mu)P_A(n) = 0, \quad \text{for} \quad n = 2, 3, ...
\end{align}
The solution of this equation is a geometrical law~[\onlinecite{morse55}, \onlinecite{schwarz75}].
By induction and using the fact that $\Sigma_{n=1}^{\infty} P_A(n) = 1$, one can show that the
solution of this equation is:
\begin{equation}
\label{eq:monoexpFit}
P_A(n) = p^{n-1}(1-p), \quad \text{where} \quad p = \frac{\lambda}{\mu}.
\end{equation}
In a log-lin scale, the stationary probability predicted by the master equation approach
is a straight line of the form :
\begin{equation}
\ln(P_A(n)) = \ln(p) n + \ln\Bigl(\frac{1-p}{p}\Bigr).
\end{equation}
It is a monoexponential distribution, as observed in Fig.~\ref{fig:pdfall} (b), obtained in MD.
The exponential distribution can be used to fit the molecular dynamics data, as shown in Fig.~\ref{fig:AonlyBB}.
The value of the dynamical parameter $p$ is, in this case, $p = 0.44$.
The master equation framework provides a dynamical interpretation of the parameter $p$ as
the ratio between the attachment and detachment rate constants.

Figure~\ref{fig:AonlyBB} also shows the probability of having a nanoaggregate containing
$n$ asphaltene molecules obtained in MD simulations for two different system sizes and the
corresponding exponential fits.
The results for a system containing $50$ asphaltene molecules and for a system containing
$5$ times less molecules agree surprisingly well, giving the values $p = 0.44$ and $p = 0.41$
respectively.
It shows that, while finite size effects are present, they are not very important
when it comes to the number of asphaltene molecules in a nanoaggregate. More specifically,
the relative probability of small nanoaggregates is accurately described even
in the small system. The probability of having larger nanoaggregates is not so well described
in the small system, as expected, but does not affect much the value of the parameter $p$
since these nanoaggregates are rare even in the large system. 
It is important to note, however, that physical properties due to large nanoaggregates or to 
nanoaggregates filling up the box in one direction, such as residual stresses, could be affected
by finite size effects.

\subsection{Aggregation dynamics}
As shown above, the stationary state predicted by the master equation agrees with the MD results on the
probability of having a nanoaggregate containing $n$ asphaltene molecules. If the master
equation~Eq.~\eqref{eq:masterEq} correctly describes the aggregation process, it should also reproduce
the aggregation dynamics. Checking this fact is the purpose of this section. 

To compare the prediction of the master equation approach and the molecular dynamics results, we
quantified the aggregation dynamics in the following way. We used the fraction $f_a$ of aggregated
asphaltene molecules versus time.
The fraction $f_a$ of aggregated asphaltene molecules is defined as the ratio of the number
of asphaltene molecules in all aggregates containing at least two asphaltene molecules and
the total number of asphaltene molecules in the system.
The fraction $f_a$ verifies:
\begin{equation}
f_a = 1-f_1,
\end{equation}
where $f_1$ is the fraction of free asphaltene molecules. 
The fraction of free asphaltene molecules is defined as
\begin{equation}
f_1 = \frac{M_{1,A}}{M_{t,A}},
\end{equation}
where $M_{1,A}$ is the number of asphaltene molecules in aggregate containing one asphaltene molecule
and $M_{t,A}$ is the total number of asphaltene molecules in the system.
The fraction $f_1$ can be expressed
in terms of the probability $P(1,t|1)$ of having an asphaltene aggregate of size $1$ at time $t$ and of
the average size of an asphaltene aggregate $\langle n \rangle_1(t)$, given that all aggregates had size 1 initially.
Indeed, we know that:
\begin{equation}
P_A(1,t|1) = \frac{N_{1, A}}{N_{t, A}},
\end{equation}
where $N_{1, A}$ is the number of aggregates containing one asphaltene molecule and $N_{t, A}$
is the total number of aggregates containing at least one asphaltene molecule,
as in Eq.~\ref{eq:PAdef}.
We also know that:
\begin{equation}
\langle n \rangle_1(t) = \frac{M_{t, A}}{N_{t, A}},
\end{equation}
where $M_{t, A}$ is again the total number of asphaltene molecules
and $N_{t, A}$ is the total number of aggregates containing at least one asphaltene molecule.
It is easy to show that $N_{1, A} = 1 \times M_{1, A}$ and finally that
\begin{equation}
f_1 = \frac{P_A(1,t|1)}{\langle n \rangle_1(t)}.
\end{equation}

To obtain the dynamics predicted by the master equation~Eq.~\eqref{eq:masterEq} with the specific boundary
condition~Eq.~\eqref{eq:BC}, a numerical implementation 
of the scheme was carried out. In the numerical implementation
the time step $\Delta t$ is a hundred times smaller than the inverse of the detachment rate constant $1/\mu$. An aggregate
can attach to a single molecule with the probability $\lambda \Delta t$ and release a single molecule with the probability $\mu \Delta t$.
An aggregate of size $1$ can attach to a molecule
but cannot release one. The total number of molecules is kept constant
in the numerical implementation of the master equation.
The total number of molecules
is chosen to be $5000$ to reduce the statistical noise.
Initially, in the numerical implementation of the master equation, all the aggregates are of size $1$.

\begin{figure}
  \includegraphics[angle=-90, scale=0.4]{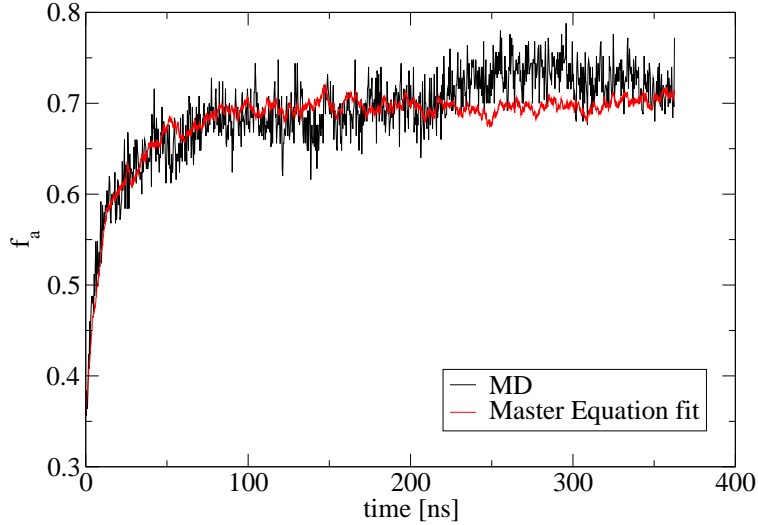}
  \caption{\label{fig:dynamics}
Time evolution of the fraction of aggregated asphaltene molecules $f_a$ in molecular
dynamics simulations and in the master equation approach.
}
\end{figure}

The fraction $f_1$ of free asphaltene molecules can be obtained through this numerical implementation. It depends
\emph{a priori} on the two rate constants $\lambda$ and $\mu$. However, the ratio $p = \lambda/\mu$
is known from the stationary state result. It leaves us with one dynamical parameter, say
$\lambda$, to fit.
The initial state in the molecular dynamics simulations is not as well defined as in the
master equation approach.
To reach the desired density, a first MD simulation where the system is compressed is performed. During
the compression period, the asphaltene molecules begin to aggregate, but the data cannot be recorded.
The data are recorded just after the compression period, in a state where small asphaltene aggregates
are already formed. The fraction of free asphaltene molecules in that state is $f_{1i} = 0.36$. It is
a second fitting parameter.
The curve predicted by the numerical implementation of the master equation
was shifted in time so that time $t=0$ corresponds to $f_{1i} = 0.36$ as in the MD simulations.
Figure~\ref{fig:dynamics} shows the fraction of aggregated asphaltene molecules versus time in the molecular
dynamics simulations and in the master equation approach. Both results
agree well, indicating that the master equation approach correctly describe the aggregation
process of asphaltene molecules.
The value of the dynamical parameter $\lambda$ is found to be $\lambda = 4.4 \cdot 10^7$ s$^{-1}$.

It is worth mentioning that the inverse rate constant $1/\lambda = 2.3 \cdot 10^{-8}$ s is much larger than
the upper limit $\tau = 5.3 \cdot 10^{-10}$ s,
needed for an asphaltene molecule to diffuse of a distance equals to
the average distance $d = 1.34$ nm between the centers of mass of $150$ aromatic molecules in a homogeneous system 
of volume $362$ nm$^3$ minus 
the intermolecular distance $d_{\text{avg}} = 4$ \AA $ $ in an
aggregate. This characteristic time is evaluated using the diffusion coefficient of a single asphaltene
molecule in the docosane solvent: $D = 2.8 \cdot 10^{-10}$ m$^2$.s$^{-1}$ and the formula
$\tau = (d-d_{\text{avg}})^2/(6D)$ for three-dimensional diffusion.
This time is an upper limit because only the distance between centers of mass is considered,
whereas molecules are extended in space and can be close to each other even if the distance
between their centers of mass is larger than $d_{\text{avg}}$.
 We can conclude from that fact
that the nanoaggregation process is not limited by diffusion.

In Fig.~\ref{fig:dynamics}, there seems to be a discrepancy between the master equation approach and the MD results
at long times. This could indicate the existence of another aggregation process, taking place at a longer time scale.
Checking carefully the existence of this second process requires further investigation.

The master equation approach related to the aggregation reaction~Eq.~\eqref{eq:reaction} is able to reproduce
both the stationary behavior and the dynamics of the asphaltene aggregation process at intermediate time scales. 
It indicates that the assumptions made to derive the master equation~Eq.~\eqref{eq:masterEq} are relevant.
The master equation approach gives a dynamical
framework to interpret the monoexponential distribution observed in MD. 
Even if the same master equation has already been used in the general field of clustering
processes~[\onlinecite{sillrenMasterEq, erdmann, mahnke}],
it is the first time, to our knowledge, that
it is applied to asphaltene aggregation in molecular dynamics simulations.

\section{Statistical mechanics approach}
\label{sec:statMech}

Section~\ref{sec:masterEq} provided a dynamical interpretation of the monoexponential behavior of the probability
of having a nanoaggregate containing $n$ asphaltene molecules
based on a master equation description on a simple aggregation reaction.
The aim of the present section is to describe a statistical mechanics model
based on the same aggregation scheme and 
consistent with the monoexponential behavior.
In this section, we will also look into possible thermodynamical
interpretations of the biexponential behavior, obtained when all aromatic molecules are counted.

\subsection{Statistical mechanics model}
\label{sec:thermoModel}
To formulate a simple statistical mechanics model linked to the aggregation reaction~Eq.~\eqref{eq:reaction},
some assumptions are needed.
We make the following common assumptions:
\begin{enumerate}
\item The nanoaggregates are linear.
\item The system is homogeneous and isotropic on average and the nanoaggregates are rigid.
\item The aggregation process occurs at a characteristic time much larger than
the equilibration of the pressure, temperature and solvent molecules.
\item The system is dilute enough to consider no interaction
between the nanoaggregates except through the aggregation reaction. In other
words, the solution of nanoaggregates of different sizes in a solvent is ideal.
\item The free energy of a nanoaggregate depends linearly on its size.
\end{enumerate}
The idea of considering a mixture of free molecules and dimers as an ideal mixture
with an aggregation reaction occurring at a larger time scale dates back to the beginning of the 20$^{\text{th}}$
century~[\onlinecite{dolezalek1908, dolezalek1910}] as is nicely explained in
the recent review~[\onlinecite{gubbins}]. 
The set of assumptions was then completed to include the case of linear aggregates
of any size and widely used to describe rodlike micelles~[\onlinecite{cates1991}],
linear polymer chains~[\onlinecite{wittmer98, wittmer2000}], and discotic liquid crystals~[\onlinecite{hurt}].
We propose in this section
our own version of the derivation applied to the case of linear asphaltene aggregates
and obtain an analytical formula for the probability of having an aggregate of a given size.
The analogy between linear assembly in discotic liquid crystals and asphaltene aggregates
was also made in Ref~[\onlinecite{aguilera2006}], which derives very similar equations to
the ones presented in this section, but do not present a successful comparison
with their coarse-grained simulations of asphaltene-resin nanoaggregates at high density.

The assumption of having an ideal mixture can be relaxed to take into account more complicated
interactions between the aggregates such as excluded volume interactions~[\onlinecite{doi, vanderSchoot}],
the gain in entropy when the chain breaks~[\onlinecite{wittmer98}] and
interactions due to the flexibility of the aggregates[\onlinecite{khokhlov,kuriabova,demichele}],
to cite only a few.
For the sake of simplicity, these interactions are neglected in the case of linear asphaltene
aggregates.
The validity of the assumptions made in this section will be discussed in Sec.~\ref{sec:assumption}.

Concurrently to these approaches expressing the probability of having an aggregate of a given size,
another theory of aggregation based on statistical thermodynamics was developed
by Wertheim~[\onlinecite{wertheim, gubbins}]. This theory considers molecules of a reference liquid,
typically a Lennard-Jones fluid, with a finite number of binding sites. The potential modeling
the interaction between the binding sites is designed to take into account steric hindrance.
The theory then counts the number of molecules with no bond on a given binding site instead of the
number of aggregates of a given size. It is a clever way to count physically meaningful graphs.
The outcome of the theory is an expression for the equilibrium pressure and the concentration of
free molecules versus the composition of the system. These predictions were successfully checked
numerically for strongly associating fluid~[\onlinecite{johnson}]. However, this theory
does not provide an expression for the probability of having an aggregate of a given size.
For this reason, it is not considered in further details in this section but is mentioned for the
sake of completeness.

In the framework of the common assumptions stated above, 
the free energy of the system can be written as~[\onlinecite{Diu}]:
\begin{equation}
\label{eq:lastExpressionF}
F = \sum\limits_n  k_B  T \Bigl( N_{n, A}(\ln(N_{n, A})-1) - N_{n, A}\ln(V)\Bigr) + N_{n, A} F_e^{(n)},
\end{equation}
where $N_{n, A}$ is the number of nanoaggregates containing $n$ asphaltene molecules,
$V$ the volume of the system, $F_e^{(n)}$ the effective free energy of an aggregate,
$k_B$ the Boltzmann constant and $T$ the temperature.
The first term in Eq.~\eqref{eq:lastExpressionF} corresponds to the free energy
of an ideal mixture of ideal gas and the second term corresponds to the energy of the aggregates.
The effective free energy $F_e^{(n)}$ is to be understood as the energy
of an asphaltene aggregate when the degrees of freedom due to solvent molecules have been integrated out.

The expression Eq.~\eqref{eq:lastExpressionF} for the free energy of the system holds for each value
of the number $N_{n, A}$ of aggregates of a given size, because a state of local equilibrium 
acting on entropy, pressure and solvent molecules is
assumed for each step along the aggregation reactions. The total equilibrium of the system depends
now only on the equilibrium of the aggregation reactions~\eqref{eq:reaction}. The condition
for chemical equilibrium of each aggregation reaction $A_n +A_1 \rightleftharpoons A_{n+1}$ is
\begin{equation}
 \mu_{n+1}-\mu_n-\mu_1 = 0,\quad \text{for } n = 1, 2, ...
\end{equation}
where $\mu_n$ is the chemical potential of an asphaltene aggregate of size $n$.
Using the definition of the chemical potential $\mu_n = \partial F/\partial N_{n, A}$,
we can obtain after some calculations
the well-known mass action law:
\begin{equation}
\label{eq:massActionLaw}
\frac{N_{n+1, A}}{N_{n, A} N_{1, A}} = \frac{K_n(T)}{V}, \quad \text{for} \quad n = 1,2,...
\end{equation}
where $K_n(T)$ is the equilibrium constant of the
reaction $A_n +A_1 \rightleftharpoons A_{n+1}$.
The equilibrium constant is given by:
\begin{equation}
\label{eq:eqCst}
K_n(T) = \exp\Bigl( -\frac{F_e^{(n+1)}-F_e^{(n)} - F_e^{(1)}}{k_BT}\Bigr).
\end{equation}
We now make use of assumption 5, stating that the free energy of a nanoaggregate
depends linearly on its size.
This assumption amounts to write the effective free energy $F_e^{(n)}$ as:
\begin{equation}
F_e^{(n)} = n F_0 + (n-1) F_e,
\end{equation}
where $F_0$ is the free energy of a single asphaltene molecule
and $F_e$ is the effective free energy between two asphaltene molecules.
Using this form in the expression of the equilibrium constant~Eq.~\eqref{eq:eqCst} gives:
\begin{equation}
K(T) = \exp\Bigl( -\frac{F_e}{k_BT}\Bigr).
\end{equation}
Thus, the equilibrium constant does not depend on the size $n$ of the considered nanoaggregate
under the set of assumptions considered.
By induction, it is now easy to show from~Eq.~\eqref{eq:massActionLaw} that
\begin{equation}
\label{eq:expressionNn}
N_{n, A} = N_{1, A}^n \Bigl( \frac{K(T)}{V}\Bigr)^{n-1}.
\end{equation}
To compare the model to the simulations results it is more useful to obtain the probability
$P_A(n)$ of having a nanoaggregate of size $n$, which is defined in Eq.\ref{eq:PAdef}
\begin{equation}
P_A(n) = \frac{N_{n,A}}{N_{t,A}},
\end{equation}
where $N_{t,A} = \sum_n N_{n,A}$ is the total number of asphaltene nanoaggregates.
To express $P_A(n)$, we make use of the conservation of the total number $M_{t, A}$ of asphaltene molecules:
\begin{equation}
\label{eq:conservation}
M_{t, A} = \sum\limits_n n N_{n, A}.
\end{equation}
Having this in mind, one can show (see Appendix~\ref{ap:paramp}) that
\begin{equation}
P_A(n) = p^{n-1}(1-p)
\end{equation}
where
\begin{equation}
\label{eq:pexpression}
p = \frac{x+1-\sqrt{2x+1}}{x},
\end{equation}
and
\begin{equation}
x = \frac{2 M_{t, A}}{V} \exp\Bigl(-\frac{F_e}{k_BT}\Bigr).
\end{equation}
We have now recovered the exponential distribution
observed in the MD simulations and Fig.~\ref{fig:pdfall} (b).
The MD simulations provide a value of the parameter $p$ characterizing the exponential distribution:
$p= 0.44$.
According to the thermodynamical interpretation~Eq.~\eqref{eq:pexpression}, it leads to the value
\begin{equation}
F_e = -4.0 \text{  }k_B T.
\end{equation}
The simple statistical mechanics model provides
an interpretation for the parameter $p$ in terms of an effective free energy $F_e$
between two asphaltene molecules.
Within the assumptions of this simple model, the effective free energy 
is the interaction energy between two asphaltene molecules when the degrees of freedom related to
solvent molecules are integrated out.

The physical meaning of free energy $F_e$ depends on its definition in the framework of the assumptions made here,
but also on the validity of these assumptions. This will be discussed in the next section.

\subsection{Physical meaning of the effective free energy and validity of the assumptions}
\label{sec:assumption}
The validity of each assumption and its consequences on the physical meaning of the free energy $F_e$
will now be listed.
\begin{enumerate}
\item \emph{Linear dependence of $F_e^{(n)}$ on the size $n$ of the nanoaggregate.}

It is very easy to check that the potential energy, and not the free energy,
of a nanoaggregate in vacuum depends linearly on its size.
The potential energy of a linear nanoaggregate of size $n$ is plotted versus $n$ for nanoaggregates in vacuum
in Fig.~\ref{fig:potEnergyA}. The origin of the energy is set arbitrarily to zero in this figure.
The nanoaggregate used to plot this figure is a linear nanoaggregate containing five asphaltene molecules
found in one simulation.
To find the potential energy of an aggregate of size $n\leq5$, only the first $n$ asphaltene
molecules in the aggregate were considered in vacuum.
This figure clearly shows that the potential energy of a nanoaggregate in vacuum is linear
and the slope $U_{\text{vacuum}}$ corresponding to the interaction energy between two asphaltene
molecules is equal to
\begin{equation}
U_{\text{vacuum}} = -87 \text{  }k_B T.
\end{equation}
In practical terms, it means
that when an asphaltene molecule is added to a nanoaggregate of size $n$, this asphaltene molecule interacts
with the energy $-87$  $k_B T$ with the molecule at the end of the nanoaggregate but do not interact with the other ones, which are further away.

\begin{figure}
  \includegraphics[angle=-90, scale=0.4]{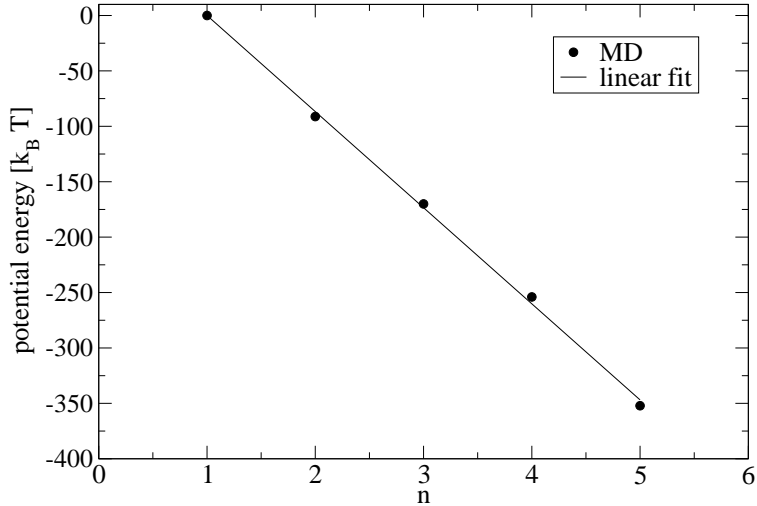}
  \caption{\label{fig:potEnergyA}
Potential energy of a linear asphaltene nanoaggregate in vacuum versus the number $n$ of asphaltene
molecules which it contains.
}
\end{figure}

The fact that the potential energy of a nanoaggregate in vacuum depends linearly on its size 
is a good indication that it might also be the case for the effective free energy $F_e^{(n)}$ of a nanoaggregate
in a solvent. In that case, the effective free energy $F_e$ corresponding to the slope of 
$F_e^{(n)}$ versus the size $n$ of the nanoaggregate is well defined.

We remind the reader that the potential energy of a nanoaggregate in vacuum 
and the effective free energy $F_e^{(n)}$ are not identical and the slopes $F_e$ and $U_{\text{vacuum}}$
are indeed quite different. 
Within the framework of the statistical model presented in Sec.~\ref{sec:thermoModel}, the two
causes of this difference are: the interaction between the asphaltene molecules and the solvent
and the entropic effects due to the effective integration of the degrees of freedom of the solvent.
To quantify the effective free energy $F_e$ properly,
a possibility is to 
set up umbrella sampling simulations controlling the distance between the center of mass
of two asphaltene molecules in a bath of docosane molecules at the same temperature and pressure as the one used in the MD simulations.
The free energy of this system can be derived in terms of the distance between the two asphaltene molecules.
The effective free energy $F_e$ would then be the difference between the free energy of such a system
when the two asphaltene molecules are far away and the free energy of the system when the two asphaltene
molecules are aligned and close.
Implementing umbrella sampling simulations is beyond the scope of this paper.
Moreover, the free energy $F_e$ found with this method might be slightly different from the value $F_e = -4.0$  $k_B T$,
found using the statistical mechanics model of Sec.~\ref{sec:thermoModel}, because of the other assumptions
made to derive this model.

\item \emph{Asphaltene nanoaggregates.}

To derive the statistical mechanics model in Sec.~\ref{sec:thermoModel}, we only considered the aggregation of asphaltene
molecules, whereas resin and resinous oil molecules are also part of the nanoaggregates.
The degrees of freedom related to the position of resin and resinous oil molecules should be integrated in an effective
way to obtain the free energy $F_e$, just as it was done for the degrees of freedom related to the solvent.
The fact that the monoexponential behavior is valid for the probability of having a nanoaggregate containing $n$ asphaltene
molecules means that this effective integration can be done.

\item \emph{Linear nanoaggregates.}

The nanoaggregates considered in the simulations are only some linear portions of bigger branched nanoaggregates. 
The existence of the branches could also produce some degrees of freedom to be integrated to obtain the effective
free energy $F_e$.

\item \emph{Dilute limit.}

The assumption stating that the solution of nanoaggregates is ideal neglects the interaction between the nanoaggregates.
Some of these interactions, for example between the nanoaggregates
$A_n$, $A_{n+1}$ and $A_1$, are later taken into account through the aggregation reactions~\eqref{eq:reaction}
corresponding to the asphaltene aggregation. But in the MD simulations, the asphaltene molecules interact in a more
complicated way. They cannot for example be too close to each other, due to short range repulsion.
This could be taken into account through excluded volume interactions. 
It is known~[\onlinecite{vanderSchoot}, \onlinecite{kuriabova}],
that taking into account excluded volume interactions preserves the monoexponential behavior. Consequently,
excluded volume interactions might play a role in the value of the effective free energy $F_e$.

\item \emph{Time scale of the aggregation reactions.}

To derive the statistical mechanics model, we assumed that the aggregation reactions occurred
on a time scale much larger than the time scale associated with the equilibrium of pressure, entropy, and solvent molecules
for a given number of each nanoaggregate.
It is very difficult to predict the effect of the relaxation of this fundamental assumption.
One way to check the assumption, however, is to compare once again the characteristic time of the aggregation
reaction $1/\lambda = 2.3 \cdot 10^{-8}$ s and the upper limit $\tau = 5.3 \cdot 10^{-10}$ s for the diffusion
of an asphaltene molecule. A factor $20$
exists between the two characteristic times which should be enough to ensure the validity of the
assumption.

\item \emph{Rigidity of the nanoaggregates and isotropy of the system.}

\label{sec:relaxRigidity}
The assumptions of rigid nanoaggregates and isotropic system can be considered together,
because if they are both relaxed, they lead to a new term in the free energy of the system~[\onlinecite{lu2004, lu2006, kuriabova}].
This term depends on a persistence length $l_p$ and reads as~[\onlinecite{khokhlov},\onlinecite{kuriabova}]
\begin{equation}
\label{eq:Fflexible}
F_{\text{flexible}} = -V\frac{2L}{3l_p}\sum\limits_n \int \frac{d\Omega_{\mathbf{u}}}{4\pi} [\rho_n(\mathbf{u})]^{1/2} \nabla[\rho_n(\mathbf{u})]^{-1/2},
\end{equation}
where
\begin{equation}
\rho_n(\mathbf{u}) = \frac{nN_n}{V}
\end{equation}
is the number density of asphaltene molecules part of a nanoaggregate of size $n$ oriented in the direction $\mathbf{u}$
with respect to some reference direction and $\Omega_{\mathbf{u}}$ is the corresponding solid angle.
This term is helpful to describe the nematic phase of liquid crystals, where long range order is seen.
The addition of this term in the free energy expression leads to a biexponential behavior~[\onlinecite{lu2004, lu2006, kuriabova}],
which is not observed
for the probability of having a linear nanoaggregates with $n$ asphaltene molecules.
Consequently, the assumption of rigid nanoaggregates and isotropic system probably holds, at least in an effective way, for 
the probability of having a linear nanoaggregate with $n$ asphaltene molecules
and does not participate in the value of the effective free energy $F_e$.
The addition of this flexible term to the free energy is potentially of interest to explain 
the biexponential behavior of the probability of having a nanoaggregate containing $n$ aromatic molecules.
\end{enumerate}

To summarize, one can say that the statistical mechanics model developed in Sec.~\ref{sec:thermoModel}
provides a thermodynamical interpretation of the monoexponential behavior based on the effective free energy
between two asphaltene molecules. 
This effective free energy should be understood
as the interaction energy between two asphaltene molecules when all the degrees of freedom related to
solvent molecules, resin and resinous oil molecules and branched nanoaggregates have been integrated out.

\subsection{Biexponential behavior}
\label{sec:biexponential}
The picture emerging from the study of the probability of having a linear nanoaggregate containing
$n$ asphaltene molecules is that asphaltene bodies interact with an effective energy through
the aggregation reaction~\eqref{eq:reaction}.
This leads to a monoexponential behavior. We know that the probability of having a linear nanoaggregate
with $n$ aromatic molecules has a different behavior. It is biexponential as shown in Fig.~\ref{fig:pdfall} (a).
This section is devoted to identifying
possible statistical mechanics explanations for this different behavior.

\subsubsection{Role of resin and resinous oil molecules}
The first explanation that comes to mind is that when all molecules are counted, different
interaction energies are involved. Each interaction energy taken separately would lead to a specific
monoexponential behavior and the combinations of several interaction energies could
lead to a bi- or multiexponential behavior.

To test this idea, we set up one-dimensional lattice Monte-Carlo simulations.
A linear lattice of $N$ sites is created. Each site can contain one asphaltene molecule, one
resin molecule, one resinous oil molecule or nothing.
One site cannot contain two molecules.
The total numbers of asphaltene molecules, resin molecules, and resinous oil molecules are constant.
A Monte-Carlo move consists in exchanging the content of two sites providing that the content is different.
This condition makes the molecules indiscernible. The system is initialised with the largest possible aggregate
where all asphaltene molecules are next to each other, then comes resin molecules and then resinous oil molecules.
Ten million ($10^7$) Monte-Carlo moves are realised using the Metropolis algorithm. 
The algorithm converges quite quickly despite its elementary implementation.
The potential energy of a nanoaggregate is calculated in the following way: when two asphaltene molecules
are next to each other the interaction $u_{A}$ between two asphaltene molecules is added, when a resinous oil molecule
is next to an asphaltene molecule or another resinous oil molecule the interaction energy $u_{RO}$ is added,
finally when a resin molecule is next to any other molecule the interaction energy $u_{R}$ is added.
As two molecules cannot be on the same site, effective excluded volume interactions are created.
The assumptions underlying the establishment of the one-dimensional lattice Monte-Carlo simulations are similar
to the ones made in Sec.~\ref{sec:thermoModel} except for excluded volume interactions:
\begin{enumerate}
\item The nanoaggregates can only be linear because the system is one-dimensional.
\item The nanoaggregates are rigid. The system is one dimensional, so isotropy is not a criterion.
\item The only energies involved are those related to the aggregation process. All potential energies
related to interaction with and within solvent molecules are averaged out. No kinetic energy is involved.
\item The system is not dilute and excluded volume interactions are taken into account.
\item The energy of a pure asphaltene nanoaggregate depends linearly on its size.
\end{enumerate}
One of the main advantages of lattice Monte-Carlo simulations compared to the analytical approach is to take into account
resin and resinous oil molecules and not only asphaltene molecules.
We checked that in the dilute limit, when only asphaltene molecules are present in the Monte-Carlo simulations,
the same monoexponential behavior with the same value for the parameter $p$ as the one predicted
by the analytical approach is recovered.

\begin{figure}
  \includegraphics[angle=-90, scale=0.2]{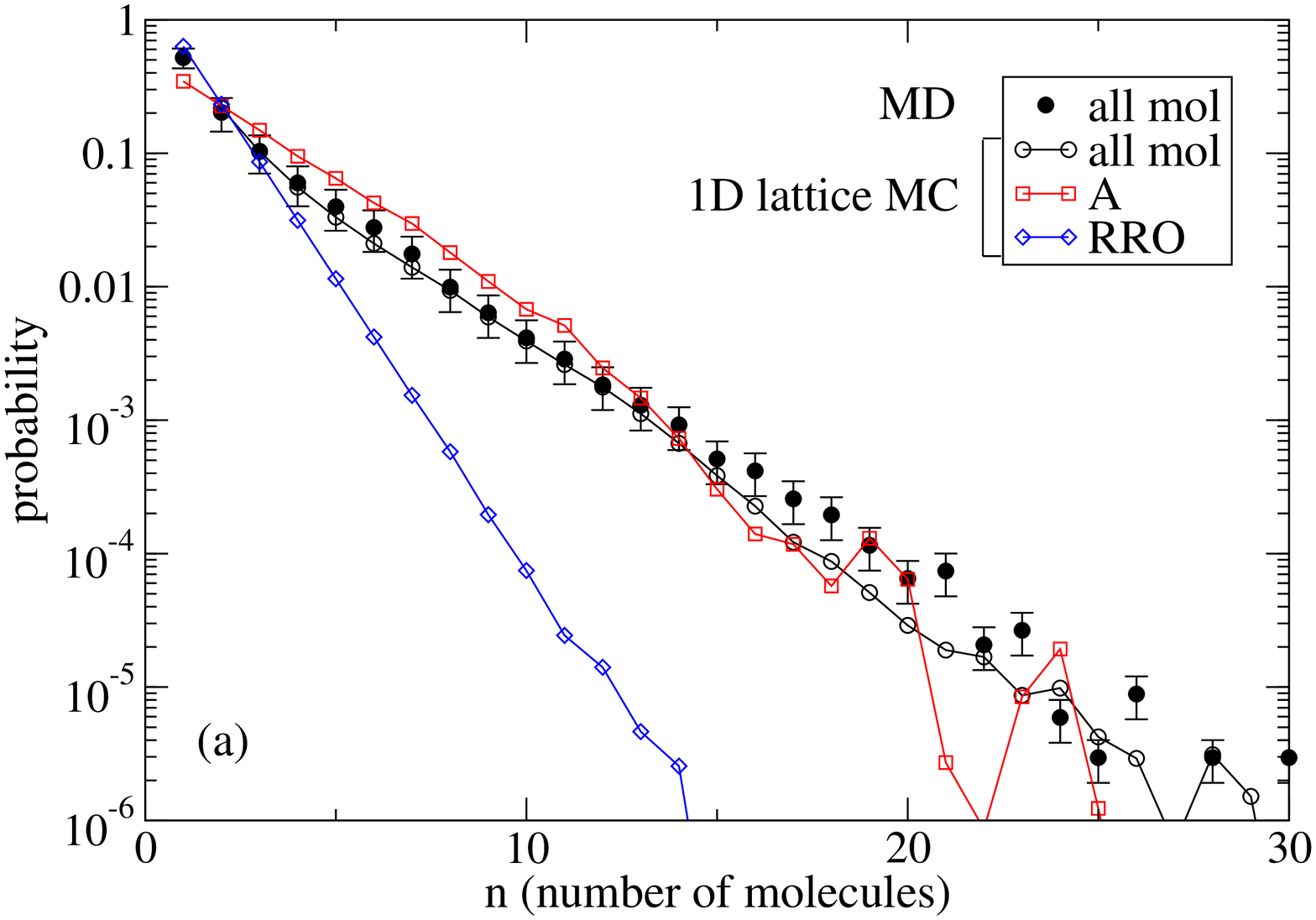} \includegraphics[angle=-90, scale=0.2]{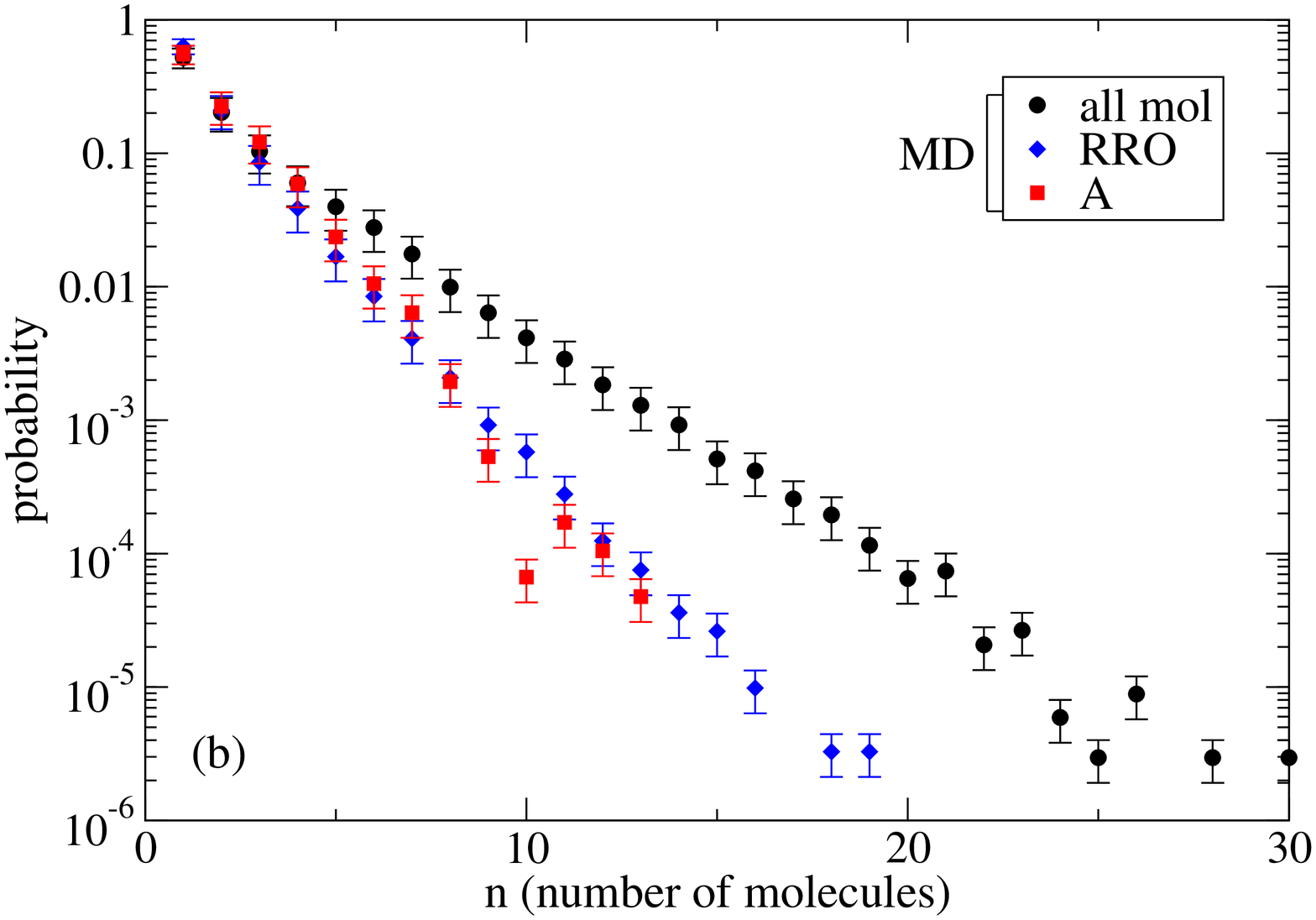}
  \includegraphics[angle=-90, scale=0.2]{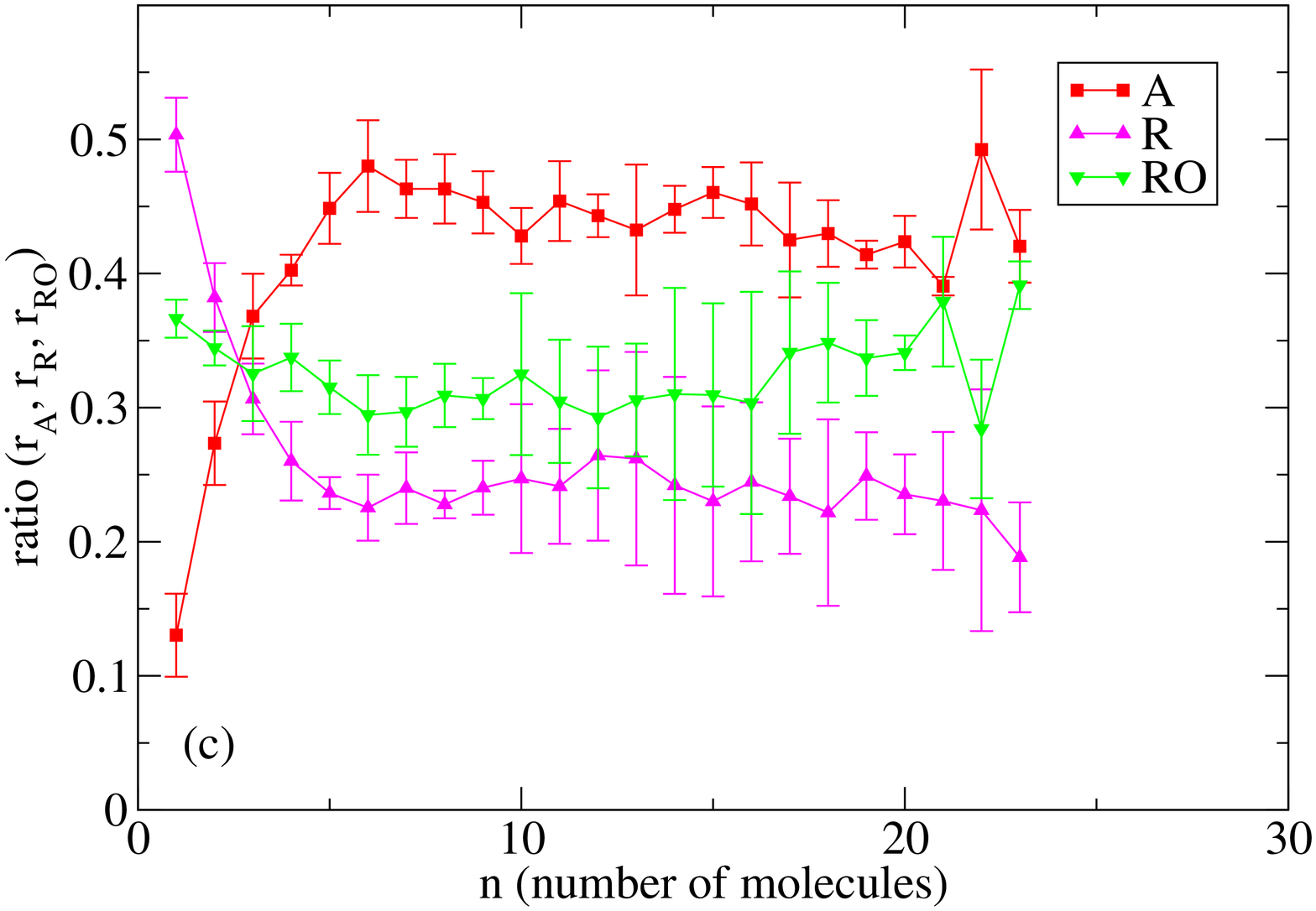}
  \caption{\label{fig:1dlatMC}
(Color online) (a) Probability of having a nanoaggregate containing $n$ aromatic molecules obtained
in molecular dynamics and in the one-dimensional lattice Monte-Carlo simulations. The probabilities of having
a nanoaggregate containing $n$ asphaltene molecules and $n$ resin or resinous oil molecules
obtained in Monte-Carlo simulations are also shown to see the difference between the two slopes.
(b) Probabilities of having a nanoaggregate containing $n$ molecules obtained
in molecular dynamics, to be compared with the results of Monte-Carlo simulations shown in (a).
(c) Ratios $r_A$,  $r_R$, and $r_{RO}$ of asphaltene,
resin, and resinous oil molecules respectively versus 
the total number $n$ of molecules in the aggregate. The error bars correspond to standard deviation
over eight independent simulations.
The largest aggregate size considered here is $23$ because all the aggregates from size $1$ to $23$
appear at least once in all simulations. Some aggregates of higher size appear only in some simulations.
}
\end{figure}

In the lattice Monte-Carlo simulations, it is possible to obtain a biexponential behavior for the
probability of having $n$ molecules in a nanoaggregate as can be seen in Fig.~\ref{fig:1dlatMC}a. 
The biexponential behavior is characterized by the presence of two straight lines with different slopes
in a log-lin scale.
The biexponential behavior occurs in the lattice Monte-Carlo simulations when the interaction energies
$u_{A}$ between asphaltene molecules on the one hand and $u_{RO}$ and $u_{R}$ with resin and resinous oil molecules
on the other hand are substantially different. For example, the choices $u_{A} = -5$ $k_B T$, $u_{RO} = -2.7$ $k_B T$ and $u_{R} = -2.3$ $k_BT$
at the same temperature and same volume as the MD simulations give a biexponential behavior very close
to the one obtained in MD in Fig.~\ref{fig:pdfall} (a). 
It is shown in Fig.~\ref{fig:1dlatMC} (a).
However, the biexponential behavior does not have the same causes as in the MD simulations.
In the Monte-Carlo simulations, the biexponential behavior is due
to the fact that pure resin or resinous oil nanoaggregates and pure asphaltene nanoaggregates are formed. There are
very few mixed nanoaggregates. Thus, the first slope in the biexponential behavior
is due to resin and resinous oil nanoaggregates and the second 
slope is due to asphaltene nanoaggregates. One consequence of this fact is that
the probability of having a nanoaggregate with $n$ resin or resinous oil molecules has a monoexponential behavior with a slope
very close to the first slope of the biexponential behavior. In the same way,
the probability of having a nanoaggregate with $n$ asphaltene molecules has a monoexponential behavior
with a slope very close to the second slope of the biexponential behavior. This can be seen in Fig.~\ref{fig:1dlatMC}(a).
On the contrary, in the MD simulations, the two slopes of the biexponential behavior do not correspond to two
different slopes in two different monoexponential behaviors. This can be seen in Fig.~\ref{fig:1dlatMC}(b). 
The existence of mixed nanoaggregates can be checked directly in the MD simulations.
Fig.~\ref{fig:1dlatMC}(c) displays the ratios $r_A$, $r_R$, and $r_{RO}$ of asphaltene,
resin, and resinous oil molecules respectively
versus the size of the aggregate. For a given molecule type $M$, the ratio $r_M$ is defined as:
\begin{equation}
r_M = \frac{n_M}{n},
\end{equation}
where $n_M$ is the number of molecules of type $M$ in the aggregate and $n$ the total number
of molecules in the aggregate. 
Fig.~\ref{fig:1dlatMC}(c) shows that there is indeed a change in the nanoaggregate composition with their size.
The ratio of asphaltene molecules increases versus the size of the aggregates until it reaches
an approximately constant value for aggregates of size $n \geq 6$.
At the same time, the ratio of resin molecules decreases versus the size of the aggregates
and reaches an approximately constant value for aggregates of size $n \geq 6$, while
the ratio of resinous oil molecules is roughly constant.
However, for any size the nanoaggregates contain all molecule types.

We can conclude that,
the biexponential behavior in MD is probably not 
only due to the difference in effective interaction energies involved.

\subsubsection{Flexible nanoaggregates}
Another possible explanation of the biexponential behavior obtained for the probability
of having a nanoaggregate containing $n$ aromatic molecules, is that it derives from the nanoaggregate flexibility.

As mentioned in Sec.~\ref{sec:assumption}, if the aggregates are flexible and the system is not isotropic,
an extra term in the free energy of the system should be considered (see~Eq.~\eqref{eq:Fflexible}).
This leads to a biexponential behavior. The existence
of the biexponential behavior can be explained qualitatively in this way: bent nanoaggregates, less stable than
straight ones, tend to be smaller and are responsible for the first slope of the biexponential behavior;
on the contrary straight nanoaggregates are larger and give rise to the second slope~[\onlinecite{kuriabova}].

\begin{figure}
  \includegraphics[scale=0.35, clip=true, trim=3cm 10cm 0cm 0cm]{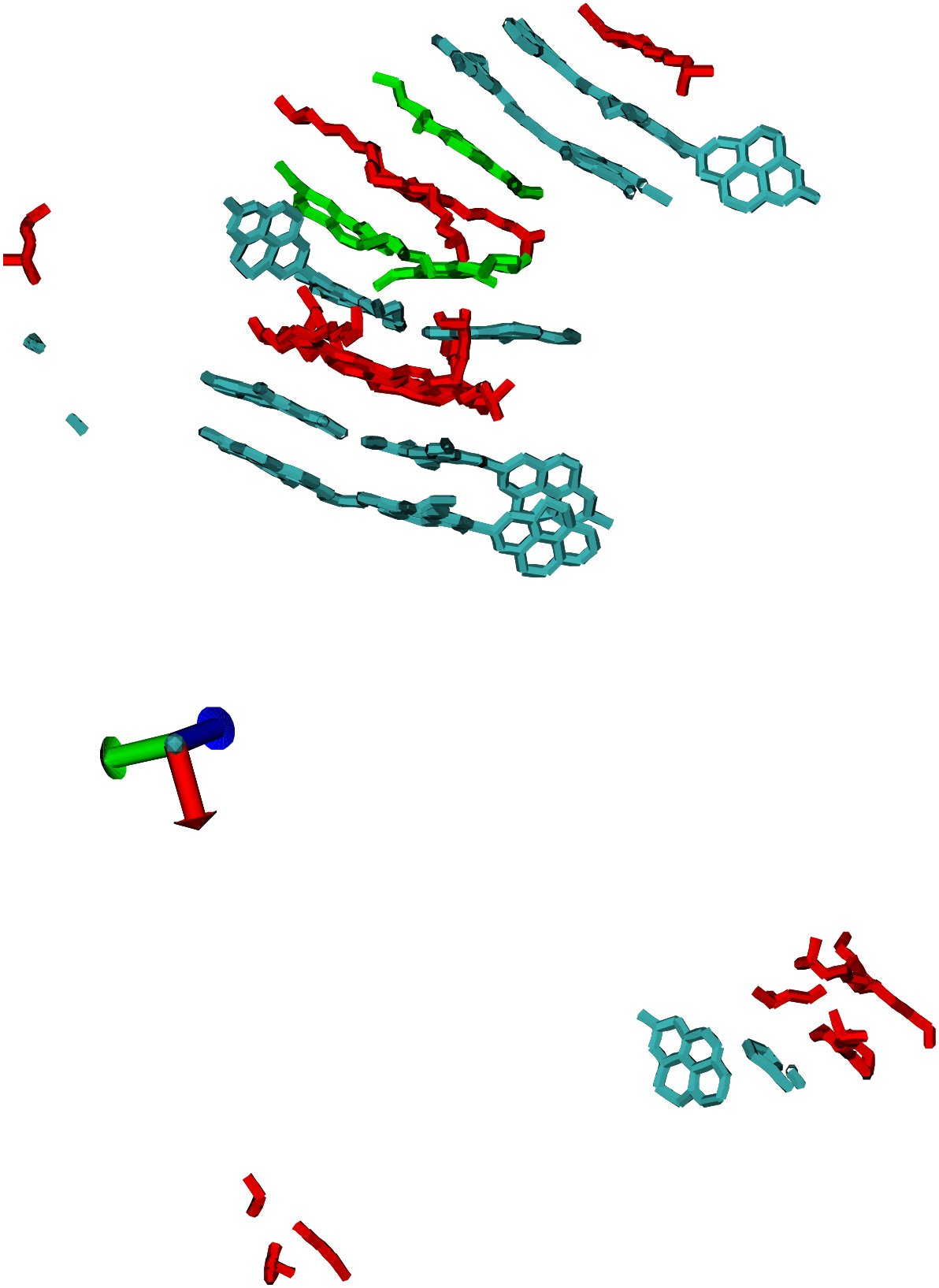}
  \caption{\label{fig:flexChain}
(Color online) Picture of a bent nanoaggregate, obtained in molecular dynamics. The color code is the same as in Fig.~\ref{fig:linearAggregate}.
}
\end{figure}

\begin{figure}
  \includegraphics[scale=0.4,angle=-90]{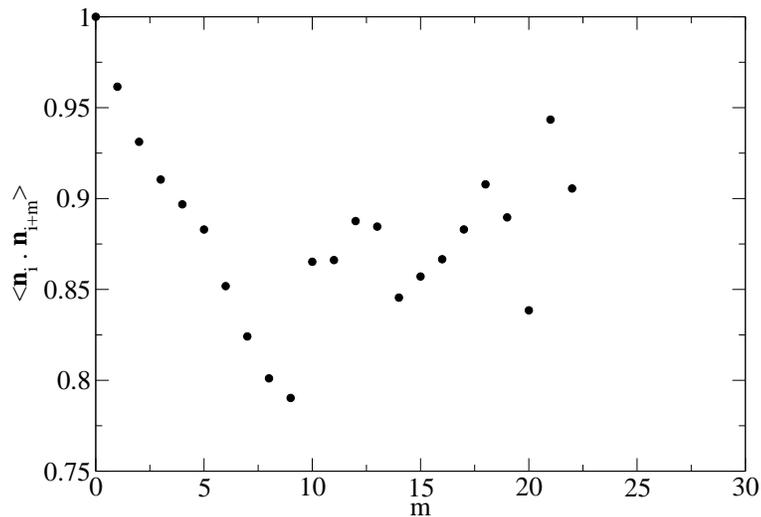}
  \caption{\label{fig:orienCorrel} Variation of the orientation correlation function
$\langle \mathbf{n}_{i} \cdot \mathbf{n}_{i+m}\rangle$ with $m$ for
linear nanoaggregates.
$\mathbf{n}_{i}$ is the unit vector
normal to the molecule $i$ in a given linear aggregate, $i+m$ stands for the $m^{\text{th}}$ neighbor
of molecule $i$ in the same linear aggregate, and $\langle \cdot \rangle$ corresponds to the average over
different nanoaggregates and over time.
}
\end{figure}

This additional term is of course well-suited to explain the biexponential behavior observed for
the probability of having a nanoaggregate with $n$ aromatic molecules. However, we are not convinced
that it is the main reason explaining the biexponential behavior, but there are indications that it might play a role.
First, some nanoaggregates are bent. A picture of a bent nanoaggregate can be seen in Fig.~\ref{fig:flexChain}(a). 
To quantify the variation of the rigidity of an aggregate with the distance inside this aggregate,
we computed the orientation correlation function
$\langle \mathbf{n}_{i} \cdot \mathbf{n}_{i+m}\rangle$, where $\mathbf{n}_{i}$ is the unit vector
normal to the molecule $i$ in a given linear aggregate, $i+m$ stands for the $m^{\text{th}}$ neighbor
of molecule $i$ in the same linear aggregate, and $\langle \cdot \rangle$ corresponds to the average over
different nanoaggregates and over time.
Fig.~\ref{fig:orienCorrel} shows the variation
of the orientation correlation function
$\langle \mathbf{n}_{i} \cdot \mathbf{n}_{i+m}\rangle$ with $m$.
The error bars corresponding to the standard deviation over eight independent
simulations are very large for this plot and are not shown for the sake of visibility.
Considering the large errors, only a qualitative discussion on the average trend is possible.
We can see that there is, on average, an initial decrease of the orientation correlation with
the number of neighbors.
This trend shows that nanoaggregates are not perfectly rigid.
For larger distances and consequently larger nanoaggregates, the
orientation correlation seems to plateau around the value $0.9$.
This last trend matches the qualitative picture of small aggregates being more bent
than larger aggregates. The change in the nanoaggregate composition observed in Fig.~\ref{fig:1dlatMC}(c)
could explain the change in the nanoaggregate rigidity with the nanoaggregate size:
aggregates containing many asphaltene molecules tend to be longer and more rigid.

Second, the system is not strictly isotropic.
The isotropy was quantified using the nematic order parameter $S$~[\onlinecite{kuriabova}].
To define the nematic order parameter, the following order tensor needs to be defined for each
nanoaggregate:
\begin{equation}
Q_{\alpha \beta} = \frac{1}{n} \sum\limits_{i=1}^{n} \Bigl( \frac{3}{2} n_{i, \alpha} n_{i, \beta} -\frac{1}{2}\delta_{\alpha \beta}\Bigr),
\end{equation}
where $n$ is the number of molecules in the considered linear nanoaggregate,
$\alpha$ and $\beta$ Cartesian coordinates,
$i$ the index of a molecule inside the aggregate,
$\mathbf{n}_{i}$ the unit vector normal to molecule $i$, and $\delta_{\alpha \beta}$ the Kronecker delta.
The nematic order parameter $S$ is the largest eigenvalue of
the order tensor $\mathbf{Q}$ averaged over different nanoaggregates and time.
The nematic order tensor is equal to $1$ in a system where all aggregates are perfectly aligned
and to $0$ in a perfectly isotropic system.
In the MD simulations, the nematic order parameter is equal to
\begin{equation}
S = 0.12 \pm 0.01
\end{equation}
where the error is the standard deviation corresponding to eight independent simulations.
The value of the nematic order parameter indicates a system closed to being isotropic but not quite.
It can be due to the fact that some linear nanoaggregates are branches of bigger aggregates.
They can be connected through asphaltene heads and the angle between an asphaltene head and
an asphaltene body is fixed by the dihedral potential and does not vary this much from one
asphaltene molecule to another~[\onlinecite{aging}].
It can also be due to steric hindrance: long nanoaggregates cannot interpenetrate each other
and consequently tend to align.
The facts that nanoaggregates are flexible and that the system is not perfectly isotropic
are consequently a plausible explanation for the biexponential behavior of the probability
of having a nanoaggregate containing a given number of aromatic molecules.

The fact that the nanoaggregates are not strictly rigid and that the system is not strictly isotropic was
also valid when we looked at the probability of having a nanoaggregate containing $n$ asphaltene molecules.
However, the fact that this probability has a monoexponential behavior means that
the additional term~Eq.~\eqref{eq:Fflexible} can be neglected in this case. It could be because the effective
persistence length is longer when only asphaltene molecules are considered.

\section{Discussion}
\label{sec:discussion}

Our bitumen model is quite simplified compared to a real bitumen. 
In this section, we will compare our results to other MD results and to experimental results.
We will also discuss the new idea that such a simplified model can bring to the field of asphaltene nanoaggregation
and the specific aspects of asphaltene nanoaggregation which are neglected in this simplified model.

Our model relies on the fact that the $\pi$-stacking interaction is the origin of the nanoaggregate formation.
In our MD simulations, the $\pi$-stacking is modeled through the Lennard-Jones potential.
Recent molecular dynamics
simulations of asphaltene molecules in toluene~[\onlinecite{jian}] reported
that the $\pi$-stacking
interaction is indeed the most important one to explain the nanoaggregate formation,
which justifies our model.

In our MD simulations, the average number of asphaltene molecules in a linear nanoaggregate is given
by the exponential distribution. 
We can compare the average and standard deviation predicted by the exponential distribution
to experimental results.
To do that, we considered a renormalized distribution of the nanoaggregate size, where
the aggregate of size $1$ are left out. 
In other words, free asphaltene molecules are not considered here as a nanoaggregate. 
The corresponding probability reads:
\begin{equation}
\label{eq:Pn>1}
P'_A(n) = p^{n-2} (1-p), \quad n \geq 2.
\end{equation}
In this case, the average number of asphaltene molecules in linear nanoaggregates
is
\begin{equation}
\langle n \rangle_A = \frac{2-p}{1-p} = 2.8.
\end{equation}
The average number of aromatic molecules in linear aggregates can be computed directly from the MD results
and is $\langle n \rangle_{\text{mol}} = 3.8$.
These results are in agreement with the general consensus that nanoaggregates contain less than $10$
molecules~[\onlinecite{mullins2011}]. 

Our simulation also offers a precise quantification of the polydispersity of the nanoaggregate size.
The standard deviation of the distribution given in Eq.~\eqref{eq:Pn>1} is:
\begin{equation}
\sigma_A = \frac{\sqrt{p}}{(1-p)} = 1.2
\end{equation}
Again, the standard deviation on the number of aromatic molecules in linear nanoaggregates can be computed
from the MD results and is $\sigma_{\text{mol}} = 2.3$.
These results are compatible with previous MD results reporting for example that:
for asphaltene nanoaggregates in toluene, the number of asphaltene molecules in the largest aggregate
varies from 2 to 18 depending on the asphaltene structure~[\onlinecite{jian}]; for asphaltene 
nanoaggregates in vacuum, the number of asphaltene molecules in any aggregate varies from 1 to 5
and the precise distribution depends both on temperature and the asphaltene structure~[\onlinecite{pacheco2003}].
On the experimental side, recent laser-based mass spectrometry experiments~[\onlinecite{wu2014}]
were able to obtain not only the average size of a nanoaggregate but also
the polydispersity.
The aggregation
numbers are found to range roughly from $3$ to $6$ or from $6$ to $8$ depending on
the bitumen chemical composition. It is also compatible with our results.

Moreover, our simulations bring the idea, already suggested in Ref.~[\onlinecite{aguilera2006}]
without a successful match to simulation results at high density, that
the simple monoexponential distribution is a typical distribution of the number
of asphaltene molecules in linear aggregates. This distribution is consequently
a good basis to model more complicated cases. One of these more complicated cases
is considered here: when all aromatic molecules are counted in the nanoaggregates
the probability of having a nanoaggregate of a given size becomes biexponential.
The biexponential distribution can be seen as a modification of the exponential distribution when the flexibity
of the aggregates and the anisotropy of the system are taken into account.

Many more specific aspects relevant for bitumen science can be considered and
are not treated here. 
For example, the effect of having branched nanoaggregates on the distribution of the nanoaggregate
size could be addressed. Furthermore, it seems that the presence of long aliphatic chains in the
asphaltene molecules modifies the typical structure of a nanoaggregate~[\onlinecite{murgich1996}] enhancing T-shaped
geometry ($\pi$-$\sigma$ interaction) and offset $\pi$-stacked geometry
($\sigma$-$\sigma$ interaction)
compared to $\pi$-$\pi$ geometry~[\onlinecite{pacheco2004}].
It would be very interesting to consider the effect of adding asphaltene and resin molecules
with long aliphatic chains in our simulations on the shape
of the nanoaggregate size distribution compared to the simple exponential distribution.

\section{Summary}
\label{sec:conclu}

In conclusion, we have shown that the aggregation reactions~\eqref{eq:reaction} give a good description
of the stability of linear asphaltene nanoaggregates as observed in molecular dynamics (MD) simulations. 
From these aggregation reactions,
we derived a master equation able to reproduce the monoexponential behavior of the stationary probability of having a nanoaggregate
containing a given number of asphaltene molecules in MD.
The parameter of the monoexponential behavior is interpreted as the ratio between the attachment
and detachment rate constants of a single asphaltene molecule to a nanoaggregate. 
The master equation approach is also able to reproduce the aggregation dynamics.
From the same aggregation reactions, we also derived a statistical mechanics model
giving a thermodynamics interpretation of the monoexponential behavior. The main parameter is then 
the effective free energy between two asphaltene molecules, when the degrees of freedom corresponding
to solvent molecules, resin, and resinous oil molecules and branched nanoaggregates are integrated out. 
Finally, a possible thermodynamic explanation for the biexponential behavior, observed for
the stationary probability of having a nanoaggregate of $n$ aromatic molecules in MD, is
the flexibility of these nanoaggregates.

To continue this work on bitumen nanoaggregate two directions are possible and equally interesting.
A first direction is to consider a simpler system without resin and resinous oil and even without
any possibility of branching. Then, the integration of the degrees of freedom related to the solvent molecules
could be done and the effective energy could be evaluated in an independent way. This
direction would lead to a quantitative understanding of the effective free energy between two asphaltene molecules.

The second and opposite direction is to add more molecule types to resemble a real bitumen.
For example asphaltene molecules without a head and with long alkyl chains could be added
and the consequences of this addition on the probability of having a nanoaggregate of a given size
investigated. Interesting MD simulations have been performed recently
using many molecules types~[\onlinecite{li2014}] and could serve as an inspiration.
Another interesting route is to quantify the evolution of the nanoaggregate
size distribution with the composition of the bitumen mixture.

\section*{Acknowledgements}
This work is sponsored by the grants 1337-00073B and 1335-00762B 
of the Danish Council for Independent Research $|$ Technology and Production
Science. It is in continuation of the Cooee project (CO$_2$ emission reduction
by exploitation of rolling resistance modeling of pavements), sponsored by
the Danish Council for Strategic Research. The centre for viscous liquids dynamics
"Glass and Time" is supported by the Danish National Research Foundation's grant DNRF61.

\appendix

\section{Expression of the parameter $p$ according to the statistical thermodynamics model}
\label{ap:paramp}
The probability $P_A(n)$ of having a nanoaggregate with $n$ asphaltene molecules is defined as:
\begin{equation}
\label{eq:defPnAgain}
P_A(n) = \frac{N_{n,A}}{N_{t,A}},
\end{equation}
where $N_{n,A}$ is the number of aggregates with $n$ asphaltene molecules and
$N_{t,A} = \sum_n N_{n,A}$ the total number of asphaltene nanoaggregates.
According to Eq.~\eqref{eq:expressionNn},
\begin{equation}
N_{n, A} = N_{1, A}^n \Bigl( \frac{K(T)}{V}\Bigr)^{n-1},
\end{equation}
so that one can express the total number of asphaltene nanoaggregates:
\begin{align}
N_{t,A} &= \sum\limits_{n=1}^{\infty} N_{1, A}^n \Bigl( \frac{K(T)}{V}\Bigr)^{n-1},\\
        &= N_{1, A} \sum\limits_{n=1}^{\infty} \Bigl(\frac{N_{1, A}K(T)}{V}\Bigr)^{n-1},\\
\label{eq:Ntot}
        &= \frac{N_{1, A}}{1- \frac{N_{1, A}K(T)}{V}}.
\end{align}
To obtain the last expression Eq.~\eqref{eq:Ntot}, we assume that the number
of asphaltene molecules is very large, so that the sum goes to infinity and that
the ratio $N_{1, A}K(T)/V$ is smaller than 1, \emph{i. e.} the volume $V$ is big enough
for the aggregates to develop given the total number of asphaltene molecules and the
equilibrium constant. 
Inserting the expression Eq.~\eqref{eq:Ntot} of the total number of asphaltene aggregates
back into Eq.~\eqref{eq:defPnAgain} gives:
\begin{equation}
P_A(n) = \Bigl(\frac{N_{1, A}K(T)}{V}\Bigr)^{n-1} \Bigl(1 - \frac{N_{1, A}K(T)}{V} \Bigr).
\end{equation}
It is a geometrical law of the form $P_A(n) = p^{n-1} (1-p)$ and the
parameter $p$ can be identified as:
\begin{equation}
p = \frac{N_{1, A}K(T)}{V}.
\end{equation}
An expression for $N_{1, A}$, the number of asphaltene aggregates of size $1$, is given by
the conservation law: $M_{t, A} = \sum\limits_n n N_{n, A}$, where
$M_{t, A}$ is the total number of asphaltene molecules. 
It leads to:
\begin{align}
M_{t, A} &= N_{1, A}\sum\limits_{n=1}^{\infty} n \Bigl( \frac{N_{1, A}K(T)}{V}\Bigr)^{n-1},\\
M_{t, A} &= \frac{N_{1, A}}{\Bigl(1-\frac{N_{1, A}K(T)}{V} \Bigr)^2},\\
\label{eq:quadratic}
0 &= N_{1, A}^2 M_{t, A} \Bigl(\frac{K(T)}{V}\Bigr)^2 - N_{1, A} \Bigl( 2 M_{t, A} \frac{K(T)}{V} +1\Bigr) + M_{t, A}.
\end{align}
The last expression Eq.~\eqref{eq:quadratic} is a quadratic equation, whose solutions are:
\begin{align}
N_{1, A}^- &= \frac{x+1-\sqrt{2x+1}}{x\times K(T)/V}\\
\text{and} \quad N_{1, A}^+ &= \frac{x+1+\sqrt{2x+1}}{x\times K(T)/V},
\end{align}
where
\begin{equation}
x = 2 M_{t, A} \frac{K(T)}{V}.
\end{equation}
The solution $N_{1, A}^-$ is the physical one, because in the limit
where there is no interaction between the asphaltene molecules and even
a large repulsion, all molecules should be in aggregates of size $1$.
In mathematical terms, it gives:
\begin{equation}
\lim_{K \rightarrow 0} N_{1, A} = M_{t, A}.
\end{equation}
Only $N_{1, A}^-$ satisfies this last equation. 
It leads finally to:
\begin{equation}
\begin{aligned}
p &= \frac{N_{1, A}K(T)}{V},\\
p &= \frac{x+1-\sqrt{2x+1}}{x},
\end{aligned}
\end{equation}
which is the same result as Eq.~\eqref{eq:pexpression}.
Another useful expression is the one giving the equilibrium constant $K(T)$
in terms of the parameter $p$:
\begin{equation}
K(T) = \frac{V}{2M_{t, A}} \frac{p+1}{(p-1)^2}.
\end{equation} 

Finally, it is noteworthy that the minimization of the free energy of
the system as given in Eq.~\eqref{eq:lastExpressionF} subject
to the conservation condition Eq.~\eqref{eq:conservation} 
leads to the same result. 
The minimization can be done using a Lagrange multiplier
to guarantee the conservation of the total number of asphaltene molecules.
In this calculation, the free energy $F_e^{(n)}$
of an aggregate should be expressed right away as
$F_e^{(n)} = nF_0 + (n-1) F_e$,
where the arbitrary
origin of the energy $ F_0$ should be equal to $F_0 = -F_e$.
In this case, $F_e$
represents the energetic penalty of having a free end.


\begin{thebibliography}{99}

\bibitem{ec140}
J. C. Petersen,
Transportation Research Circular E-C140, Transportation Research Board (2009).

\bibitem{shrp368}
J. F. Branthaver, J. C. Petersen, R. E. Robertson, J. J. Duvall, S. S. Kim, P. M. Harnsberger, T. Mill, E. K. Ensley, F. A. Barbour, and J. F. Schabron,
Tech. Rep. SHRP-A-368, Strategic Highway Research Program (1993).

\bibitem{herrington}
P. R. Herrington,
Petrol. Sci. Technol. \textbf{16}, 743 (1998).

\bibitem{lu}
X. Lu and U. Isacsson,
Construction and Building Materials \textbf{16}, 15 (2002).

\bibitem{petersen74}
J. C. Petersen, F. A. Barbour, and S. M. Dorrence,
Proc. Association of Asphalt Paving Technologists \textbf{43}, 162 (1974).

\bibitem{yen}
T. F. Yen, J. G. Erdman, and S. S. Pollack,
Anal. Chem. \textbf{33}, 1587 (1961). 

\bibitem{mullins2012}
O. C. Mullins, H.~Sabbah, J.~Eyssautier, A.~E. Pomerantz, L.~Barr\'e, A.~B.
  Andrews, Y.~Ruiz-Morales, F.~Mostowfi, R.~McFarlane, L.~Goual, R.~Lepkowicz,
  T.~Cooper, J.~Orbulescu, R.~M. Leblanc, J.~Edwards, and R.~N. Zare,
Energy \& Fuels \textbf{26}, 3986 (2012).

\bibitem{aging}
C. A. Lemarchand, T. B. Schr\o der, J. C. Dyre, and J. S. Hansen,
J. Chem. Phys. \textbf{139}, 124506 (2013).

\bibitem{andreatta}
G. Andreatta, N. Bostrom, and O. C. Mullins,
Langmuir \textbf{21}, 2728 (2005).

\bibitem{goual}
L. Goual, M. Sedghi, H. Zeng, F. Mostowfi, R. McFarlane, and O. C. Mullins,
Fuel \textbf{90}, 2480 (2011).

\bibitem{eyssautier2011}
J. Eyssautier, P. Levitz, D. Espinat, J. Jestin, J. Gummel, I. Grillo, and L. Barr\'e,
J. Phys. Chem. B \textbf{115}, 6827 (2011).

\bibitem{murgich1996}
J. Murgich, J. Rodr\'iguez M., and Y. Aray,
Energy \& Fuels \textbf{10}, 68 (1996).

\bibitem{pacheco2003}
J. H. Pacheco-S\'anchez, I. P. Zaragoza, and J. M. Mart\'inez-Magad\'an,
Energy \& Fuels \textbf{17}, 1346 (2003).

\bibitem{zhang2007}
L. Zhang and M. L. Greenfield,
Energy \& Fuels \textbf{21}, 1712 (2007).

\bibitem{headen2011}
T. F. Headen and E. S. Boek,
Energy \& Fuels \textbf{25}, 503 (2011).

\bibitem{aguilera2006}
B. Aguilera-Mercado, C. Herdes, J. Murgich, and E. A. M\"{u}ller,
Energy \& Fuels \textbf{20}, 327 (2006).

\bibitem{us}
J. S Hansen, C. A. Lemarchand, E. Nielsen, J. C. Dyre, and T. Schr\o der,
J. Chem. Phys. \textbf{138}, 094508 (2013).

\bibitem{cooee}
CO2 emission reduction by exploitation of
rolling resistance modelling of pavements,
\url{http://www.cooee-co2.dk/}.

\bibitem{rumd}
N. Bailey, L. B\o hling, J: S. Hansen, T. Ingebrigsten, H. Larsen, C. A. Lemarchand, U. R. Pedersen, T. Schr\o der, and A. A. Veldhorst
\url{http://rumd.org}.

\bibitem{SARA}
ASTM. 1995,
\textit{Annual Book of Standards} (American Society for Testing and Materials, Philadelphia, 1995); method D-2007.

\bibitem{morse55}
P. M. Morse,
Oper. Res. \textbf{3}, 255 (1955) http://dx.doi.org/10.1287/opre.3.3.255.

\bibitem{schwarz75}
M. Schwarz, Jr and D. Poland,
J. Chem. Phys. \textbf{63}, 557 (1975).

\bibitem{sillrenMasterEq}
P. Sillr\'en. J. Bielecki, J. Mattsson, L. B\"{o}rjesson, and A. Matic,
J. Chem. Phys. \textbf{136}, 094514 (2012).

\bibitem{thesisPer}
Per Sillr\'en,
\textit{Trees, Queues and Alcohols"},
Ph.D. thesis (Chalmers University of Technology, 2013).

\bibitem{erdmann}
T. Erdmann and U. S. Schwarz,
Phys. Rev. Lett. \textbf{92}, 108102 (2004).

\bibitem{mahnke}
R. Mahnke and N. Pieret,
Phys. Rev. E \textbf{56}, 2666 (1997).

\bibitem{dolezalek1908}
F. Dolezalek,
Z. Phys. Chem. \textbf{64}, 727 (1908).

\bibitem{dolezalek1910}
F. Dolezalek,
Z. Phys. Chem. \textbf{71}, 191 (1910).

\bibitem{gubbins}
K. E. Gubbins,
Mol. Phys. \textbf{111}, 3666 (2013).

\bibitem{cates1991}
M. E. Cates, C. M. Marques, and J.-P. Bouchaud,
J. Chem. Phys. \textbf{94}, 8529 (1991).

\bibitem{wittmer98}
J. P. Wittmer, A. Milchev, and M. E. Cates,
J. Chem. Phys. \textbf{109}, 834 (1998).

\bibitem{wittmer2000}
J. P. Wittmer, P. van der Schoot, A. Milchev, and J. L. Barrat,
J. Chem. Phys. \textbf{113}, 6992 (2000).

\bibitem{hurt}
R. H. Hurt and Y. Hu,
Carbon \textbf{37}, 281 (1999).

\bibitem{doi}
M. Doi and S. F. Edwards,
\textit{The theory of polymer dynamics} (Clarendon, Oxford, 1986).

\bibitem{vanderSchoot}
P. van der Schoot and M. E. Cates,
Langmuir \textbf{10}, 670 (1994).

\bibitem{khokhlov}
A. R. Khokhlov and A. N. Semenov,
Phys. A  \textbf{112}, 605 (1982).

\bibitem{kuriabova}
T. Kuriabova, M. D. Betterton, and M. A. Glaser,
J. Mater. Chem. \textbf{20}, 10366 (2010).

\bibitem{demichele}
C. De Michele, T. Bellini, and F. Sciortino,
Macromolecules \textbf{45}, 1090 (2012).

\bibitem{wertheim}
M. S. Wertheim,
J. Stat. Phys. \textbf{35}, 19 (1984);
ibid.  \textbf{35}, 35 (1984);
ibid.  \textbf{42}, 459 (1986);
ibid.  \textbf{42}, 477 (1986).

\bibitem{johnson}
J. K. Johnson and K. E. Gubbins,
Mol. Phys.  \textbf{77}, 1033 (1992).

\bibitem{Diu}
B. Diu, C. Guthmann, D. Lederer, and B. Roulet,
\textit{\'El\'ements de physique statistique} (Hermann, Paris, 1989).

\bibitem{lu2004}
X. L\"{u} and J. T. Kindt,
J. Chem. Phys. \textbf{120}, 10328 (2004).

\bibitem{lu2006}
X. L\"{u} and J. T. Kindt,
J. Chem. Phys. \textbf{125}, 054909 (2006).

\bibitem{wiehe}
I. A. Wiehe and K. S. Liang,
Fluid Phase Equil. \textbf{117}, 201 (1996).

\bibitem{mullins2011}
O. C. Mullins,
Annu. Rev. Anal. Chem. \textbf{4}, 393 (2011).

\bibitem{jian}
C. Jian, T. Tang, and S. Bhattacharjee,
Energy \& Fuels \textbf{28}, 3604 (2014).

\bibitem{wu2014}
Q. Wu, A. E. Pomerantz, O. C. Mullins, and R. N. Zare,
Energy \& Fuels \textbf{28}, 475 (2014).

\bibitem{pacheco2004}
J. H. Pacheco-S\'anchez, F. \'Alvarez-Ram\'irez, and J. M. Mart\'inez-Magad\'an,
Energy \& Fuels \textbf{18}, 1676 (2004).

\bibitem{li2014}
D. D. Li and M. L. Greenfield,
Fuel \textbf{115}, 347 (2014).










\end{thebibliography}
\end{document}